\documentclass[preprint,11pt]{elsarticle}

\usepackage{amsmath}
\usepackage{amsthm}
\usepackage{amsfonts}
\usepackage{multirow}
\usepackage{bigdelim}
\usepackage{hyperref}
\usepackage{cleveref}
\usepackage{booktabs}
\usepackage{placeins}
\usepackage{pstricks}    
\usepackage{graphicx} 
\newcommand{\ts}[2]{\ensuremath{\mathrm{#1}_{\mathrm{#2}}}} 
\newcommand{\mv}[1]{\overline{\textbf{#1}}}
\newcommand{\cv}[1]{\widetilde{\textbf{#1}}}
\newcommand{\fv}[1]{\textbf{#1}^{\prime}}

\journal{European Journal of Mechanics B/Fluids}

\usepackage{color}

\begin{document}

\begin{frontmatter}



\title{\textcolor{black}{An assessment of turbulence models for linear hydrodynamic stability analysis of strongly swirling jets}}


\author{Lothar Rukes, Christian Oliver Paschereit, Kilian Oberleithner}

\address{Hermann-F\"ottinger-Institut, TU-Berlin, M\"uller-Breslau-Strasse 8, 10623 Berlin, Germany}

\begin{abstract}
Linear stability analysis has proven to be a useful tool in the analysis of dominant coherent structures, such as the von K\'{a}rm\'{a}n vortex street and the global spiral mode associated with the vortex breakdown of swirling jets. In recent years, linear stability analysis has been applied successfully to turbulent time-mean flows, instead of laminar base-flows, \textcolor{black}{which requires turbulent models that account for the interaction of the turbulent field with the coherent structures. To retain the stability equations of laminar flows, the Boussinesq approximation with a spatially nonuniform but isotropic eddy viscosity is typically employed.
In this work we assess the applicability of this concept to turbulent strongly swirling jets, a class of flows that is particularly unsuited for isotropic eddy viscosity models. Indeed we find that unsteady RANS simulations only match with experiments with a Reynolds stress model that accounts for an anisotropic eddy viscosity. However, linear stability analysis of the mean flow is shown to accurately predict the global mode growth rate and frequency if the employed isotropic eddy viscosity represents a least-squares approximation of the anisotropic eddy viscosity. 
Viscosities derived from the $k-\epsilon$ model did not achieve a good prediction of the mean flow nor did they allow for accurate stability calculations. We conclude from this study that linear stability analysis can be accurate for flows with strongly anisotropic turbulent viscosity and the capability of the Boussinesq approximation in terms of URANS-based mean flow prediction is not a prerequisite.}
\end{abstract}

\begin{keyword}
Local linear stability analysis \sep eddy viscosity modeling \sep coherent structures \sep PIV \sep URANS



\end{keyword}

\end{frontmatter}


\section{Introduction}
\label{sec:Intro}
\textcolor{black}{
Linear stability analysis (LSA) was originally derived to investigate the growth of infinitesimal perturbations on stationary laminar base flows. Predictions based on the linear concept become inaccurate once the perturbation reaches finite size and the nonlinear terms become relevant.
However, these nonlinear terms are less relevant than one would expect when considering simple dynamical systems. 
The wake of a circular cylinder \cite{Lugo.2014,Noack.2003,Barkley.2006}, the forced turbulent mixing layer \cite{Gaster.1985,Cohen.1994,Reau.2002b} and the forced turbulent wake \cite{Marasli.1994} are examples of flows, where a periodic disturbance wave grows on an unstable flow. The growth of the disturbance induces modifications of the mean flow that increase, up to the point where the mean flow becomes neutrally stable and perturbations stop growing and saturate \cite{Lugo.2014}. The presence of the organized wave in turbulent flows leads to the formation of coherent Reynolds stresses that interact nonlinearly with the instantaneous flow and lead to a modification of the mean flow \cite{Marasli.1994,Lugo.2014}. Hence, a linear analysis of the mean flow takes the nonlinear interaction between the coherent structure and the flow implicitly into account.\\ }   

The accuracy of mean flow LSA was demonstrated on a number of flow configurations. Pier \cite{Pier.2002} investigated the capability of stability analysis to predict the limit cycle frequency of the global mode in the cylinder wake. He showed that predictions based on the base flow were only satisfactorily in the vicinity of the critical Reynolds number and deteriorated with increasing Reynolds number. Only the frequency predictions based on an analysis of the time-mean flow matched the frequencies measured in a direct numerical simulation. Other studies on wake flows also reported a very good agreement between the global mode frequency predicted by linear stability analysis of the time-mean flow and measured frequencies \cite{Barkley.2006,Hammond.1997,Lugo.2014,LEONTINI.2010,Khor.2008,Thiria.2007,Thiria.2015}. \textcolor{black}{The flow is found to be marginally stable, with the growth rate of the global mode approximately equal to zero \cite{Barkley.2006,Lugo.2014,Thiria.2015}. Turton et al. \cite{Turton.2015} give a general demonstration that a very good approximation of the frequency and a zero growth rate of the perturbation can be derived from an analysis of the time mean field, if the perturbation is quasi-monochromatic. This scenario applies to the global mode of swirling jets that feature a clear peak in the temporal spectrum and further substantiates LSA of the time mean flow (see e.g. fig.~3 in ref.~\cite{Oberleithner.2011b}).}    

The global mode of swirling jets undergoing vortex breakdown was considered by Oberleithner et al. \cite{Oberleithner.2011b,Oberleithner.2015c}. These authors report a good agreement between the spatial shape of the global mode derived from local linear stability analysis (LSA) of the time-mean flow and Proper Orthogonal Decomposition of the underlying measurement \cite{Oberleithner.2011b}. In addition, an excellent agreement of the measured and predicted frequency of the global mode was found \cite{Oberleithner.2015c}. The local LSA of the time-mean velocity field of a forced jet is considered by Oberleithner et al. \cite{Oberleithner.2014b}. This work finds that the spatial growth rate, phase velocity, phase and amplitude of the excited coherent structures are accurately captured during the entire growth and most of the decay phase. It should be noted that for all the mentioned studies, the dynamical system was represented by a single periodic structure (with higher harmonics) and nonlinear wave-wave interactions were small. This is the framework where mean flow LSA is trustworthy as the mean-coherent interactions are the leading nonlinearities. Counter examples are subharmonic resonance in the shear layer \cite{Paschereit.1995}, the cavity flow \cite{Sipp.2007}, and forced oscillator flows \cite{Terhaar.2015a}. 

The obvious advantage of mean flow LSA is the application to turbulent flows at high Reynolds numbers. A conceptual framework for LSA of turbulent flows is given by the triple decomposition of Reynolds \& Hussain \cite{Reynolds.1972}. The velocity is decomposed into a time-mean, a coherent and a purely stochastic part. The equations to be solved are the governing equations of the coherent fluctuations linearized around the mean flow. These equations are not closed due to additional terms that represent the coherent-turbulent interactions, which requires appropriate turbulence models. There are currently two possibilities to approach the closure problem. Crouch et al. \cite{Crouch.2007} and Meliga et al. \cite{Meliga.2012b} based their analysis on the linearized Reynolds-averaged Navier-Stokes equations (RANS), including a linearized turbulence model. Another approach is to use the eddy viscosity concept to derive a scalar representation of small scale turbulence from measured or simulated data sets. The eddy viscosity is added to the molecular viscosity and LSA is carried out based on an effective viscosity. This approach has been applied successfully to the analysis of swirling wakes \cite{Viola.2014}, swirling jets undergoing vortex breakdown \cite{Oberleithner.2011b,Oberleithner.2015c}, channel flows \cite{Kitsios.2010b}, airfoils \cite{Kitsios.2011} and turbulent wakes \cite{Marasli.1994}.

In this work, we adopt the second approach\textcolor{black}{, as it can be used to} derive an eddy viscosity from experimental and numerical data. 


This study focuses on highly turbulent swirling jets undergoing vortex breakdown. The turbulence structure is very complex in the near field of these flows and constant eddy viscosity models or simple mixing length models are known to suffer due to the strong anisotropy of the turbulent stresses. 
On the other hand, these flows feature a clear dominant coherent structure that arises from a global hydrodynamic instability\cite{Oberleithner.2011b}. 
This mode arises due to hydrodynamic feedback associated with the vortex breakdown bubble. It takes the form of a single helix, setting the entire flow into a precessing motion. This structure is frequently observed in swirl-stabilized combustion where it is termed ``precessing vortex core" \cite{Syred.1997,Stohr.2012}. Its analysis and prediction is ongoing research \cite{Stohr.2012,Oberleithner.2015c}. The complex turbulent structures and the clear coherent dynamics of these flows render them suitable as a benchmark for turbulent LSA.

We assess in this study the ability of the $k-\epsilon$ mixing length model and a model based on a least-squares fit of the entire Reynolds stress tensor to account for the turbulence structure of swirling jets. The eddy viscosity is derived from measurements of two swirling jet configurations. The first features a generic swirling jet, typical of fundamental studies and the second configuration is similar to setups used in swirl stabilized combustion research. The measurements of the first configuration are complemented by unsteady RANS (URANS) simulations that provide a consistency check of the results obtained from experiments. In detail, this study addresses the following open questions:
\begin{itemize}
\item
\textit{\textcolor{black}{What is the best strategy for turbulence modeling in LSA to predict the frequency and growth rate of the global mode in swirling jets undergoing vortex breakdown?}}
\item
\textit{How do different modeling strategies change the predictions of local LSA?}
\item
\textit{Can URANS computations form a consistent basis for LSA? \textcolor{black}{Is the potential failure of the Boussinesq approximation for URANS-based simulations a show stopper for the LSA?} }
\end{itemize}  
The experimental and numerical setup used in this study is presented in \cref{sec:pivSetup} and \cref{sec:cfdSetup}, respectively. Our data analysis strategy, with particular emphasis on local LSA and the computation of the turbulent viscosity from experimental data is outlined in \cref{sec:dataAnalysis}. The time-mean velocity and eddy viscosity fields are introduced in \cref{sec:results}, together with a detailed discussion of the stability analysis results. The section terminates with the presentation of the URANS-based LSA results. A detailed comparison of the URANS simulations with the PIV measurements is given in the appendix \cref{sec:cfdVali}.
\section{The experimental setup}
\label{sec:pivSetup}
\subsection{The swirling jet facility}
\begin{figure}
\centering
\includegraphics[width= 0.75\textwidth]{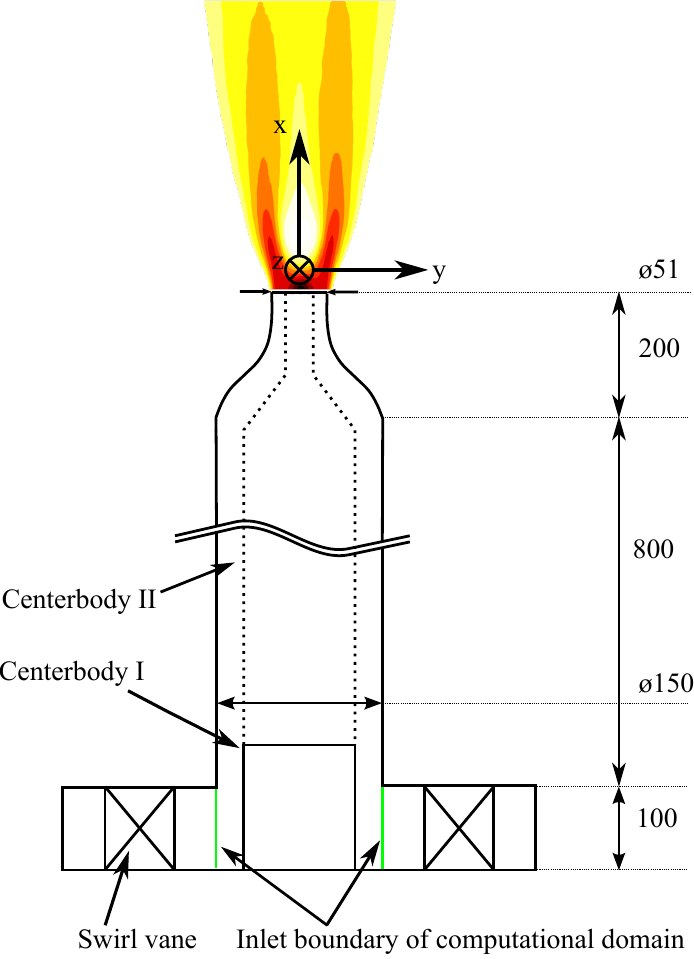}
\caption{The experimental facility. All dimensions are in millimeters and not drawn to scale. The inlet boundary of the computational domain is indicated by the green lines. Two different centerbodies are indicated by the solid and dotted black line.}
\label{fig:expFacility} 
\end{figure}

The swirling jet facility used for the experimental study is presented in \cref{fig:expFacility}. The swirling jet was generated by passing fluid from a pressurized air supply via a mass flow controller past elven swirl vanes. These vanes can be adjusted in angle simultaneously, allowing for the adjustment of different swirl intensities. The swirling flow is then guided through a pipe, followed by a nozzle and exits into the unconfined ambient. The nozzle exit diameter is 51 mm. The Reynolds number of the jet is adjusted by the mass flow rate of air coming from the pressurized air supply.

This study presents two swirling jet configurations. Both share the same Reynolds number and the same swirl angle, but they differ by centerbodies mounted onto the bottom plate of the swirler. The first has a total length of 20\% of the pipe and is indicated by the solid line in \cref{fig:expFacility}. The top fifth of this centerbody is filled with a honeycomb element to condition the flow \cite{Rukes.2015}. This configuration produces a generic swirling jet that is typical of fundamental studies on swirling jets \cite{Chigier.1965,Panda.1994,Billant.1998,Ruith.2003b,Liang.2005,Oberleithner.2011b,Oberleithner.2012,Semaan.2013,Rukes.2015}. Rukes et al. \cite{Rukes.2015} can be consulted for full details on the effect of the centerbody on the flow. The centerbody of the second configuration extends the entire length of the experimental setup and into the nozzle. This configuration is related to combustion applications, where such a centerbody configuration is often used in laboratory setups \cite{Syred.1997,Selle.2004,Stohr.2012,Oberleithner.2015c}.\\
\Cref{fig:expFacility} introduces a Cartesian coordinate system. The origin of the coordinate system is placed on the jet axis in the nozzle exit plane. The $x$-axis is oriented in the flow direction, the $y$-axis in the cross flow direction and the $z$-axis in the out-of-plane direction. Additionally, a polar coordinate system is introduced with the same origin. The $r$-axis is aligned with the $y$-axis of the Cartesian system at 0 degrees of revolution, $\theta$ is counted mathematically positive and $x$ points in the streamwise direction. In the Cartesian system, the velocity vector $\textbf{v}$ has the components $v_{x}$, $v_{y}$ and $v_{z}$. In the polar system, the velocity components read $v_{r}$, $v_{\theta}$ and $v_{x}$. \\
\Cref{fig:expFacility} also introduces the boundaries of the computational domain that was used for the URANS calculations. The computational domain included the pipe, the nozzle and the unconfined flow domain, but not the swirler. The inlet of the truncated domain is marked by the green lines in \cref{fig:expFacility}. More details on the computational setup are provided in \cref{sec:cfdSetup}.
 
\subsection{Characteristic numbers}
The swirling jets are characterized by the swirl number
\begin{equation}
\mathrm{S} = \frac{v_{\theta \mathrm{bulk}}}{v_{r \mathrm{bulk}}},
\end{equation}
with $v_{\theta \mathrm{bulk}}$ the azimuthal bulk velocity and $v_{r \mathrm{bulk}}$ the radial bulk velocity, each derived at the inlet plane of the truncated domain, see \cref{fig:expFacility}. Note that these velocities were not measured, but derived theoretically from the mass flow rate and the swirl angle. The Reynolds number is given as
\begin{equation}
\mathrm{Re} = \frac{\ts{v}{bulk} \mathrm{D}}{\nu}.
\end{equation}
\ts{v}{bulk} denotes the axial bulk velocity in the nozzle exit plane, D the nozzle diameter and $\nu$ the kinematic viscosity of air. 
The two configurations that are considered in this study, together with their experimental parameters are listed in \cref{tab:ExpPara}. Note that the centerbody of C2 extends into the nozzle and blocks 25\% of the nozzle area. Hence, the bulk velocity in the nozzle exit plane is larger. 
\begin{table}
\centering
\begin{tabular}{l c c c c}
\toprule
& $\ts{v}{bulk}$ [m/s] &  Re & S \\
\midrule
C1 & 5.8 & 20000 & 0.73  \\
C2 & 7.6 & 26000 & 0.73  \\
\bottomrule
\end{tabular}
\caption{Experimental parameters.\label{tab:ExpPara}}  
\end{table}

\subsection{Data acquisition}
Experimental data were recorded with a stereo PIV system, consisting of two PCO 2000 cameras, with a resolution of 2048x2048 pixel and a Quantel Twins BSL 200 laser, capable of emitting 170 mJ energy per pulse. Data were acquired with this system at a rate of 6 Hz. \textcolor{black}{The laser light sheet was aligned with the $x$-$y$ plane.} The double images were processed using the commercial software PIVview (PIVtec GmbH) using standard digital PIV processing \cite{Willert.1991}. The data analysis employed iterative multigrid interrogation with image deformation \cite{Scarano.2002}. The final size of the interrogation was 32x32 pixel with an overlap of 50\%. Errors in the laser sheet alignment were minimized by the use of corrected mapping functions. The initial datum calibration marks were back projected onto the measurement plane by an optimized Tsai camera model \cite{Soloff.1997}.\\
The temporal dynamic of the global mode was assessed with a Laser Doppler Anemometer measurement in the case of C1. The measurement was taken at $x/D = 0.27$, $y/D = 0.1$ and consisted of approximately 400000 samples. The frequency of the global mode for C2 was determined from the time resolved signal of eight pressure sensors (First Sensor HDO series, 10 mbar range) that were distributed circumferentially around the nozzle lip. 

\section{The URANS setup}
\label{sec:cfdSetup}
Immense efforts have been undertaken in the CFD community to develop and improve turbulence models that are based on the turbulent viscosity to close the Reynolds-averaged equations. We consider it worthwhile to attempt a URANS based simulation to retrieve the turbulent viscosity from the simulation and to use it in the LSA. We focus our attention here on C1, because the shorter centerbody allows for the formation of a strongly concentrated vortex core upstream of the nozzle \cite{Rukes.2015}, which provides a more challenging benchmark for the ability of URANS computations.

\subsection{Numerical methodology}
All results discussed in this work were obtained with the open source CFD package OpenFOAM (version 2.3.0) \cite{Weller.1998}. This software implements the finite volume method for discretizing the incompressible Navier-Stokes and continuity equations. Second order accurate schemes were used for the temporal discretization and the discretization of terms involving the velocity. First-order upwind discretization was used for turbulence quantities. The discretized equations were solved on hexahedral meshes. The URANS simulations used the PISO-algorithm \citep{Issa.1986} for the pressure-velocity coupling. The time step was chosen such that the Courant number, $\mathrm{Co} = \delta_{t} |\textbf{v}|/\delta_{x}$, had a value of 0.7 at most. $\delta_{t}$ denotes the time step and $\delta_{x}$ the local grid size. This choice of the time step size ensured solver stability.

\subsection{The computational domain}
\begin{figure}
\centering
\includegraphics[width= 0.75\textwidth]{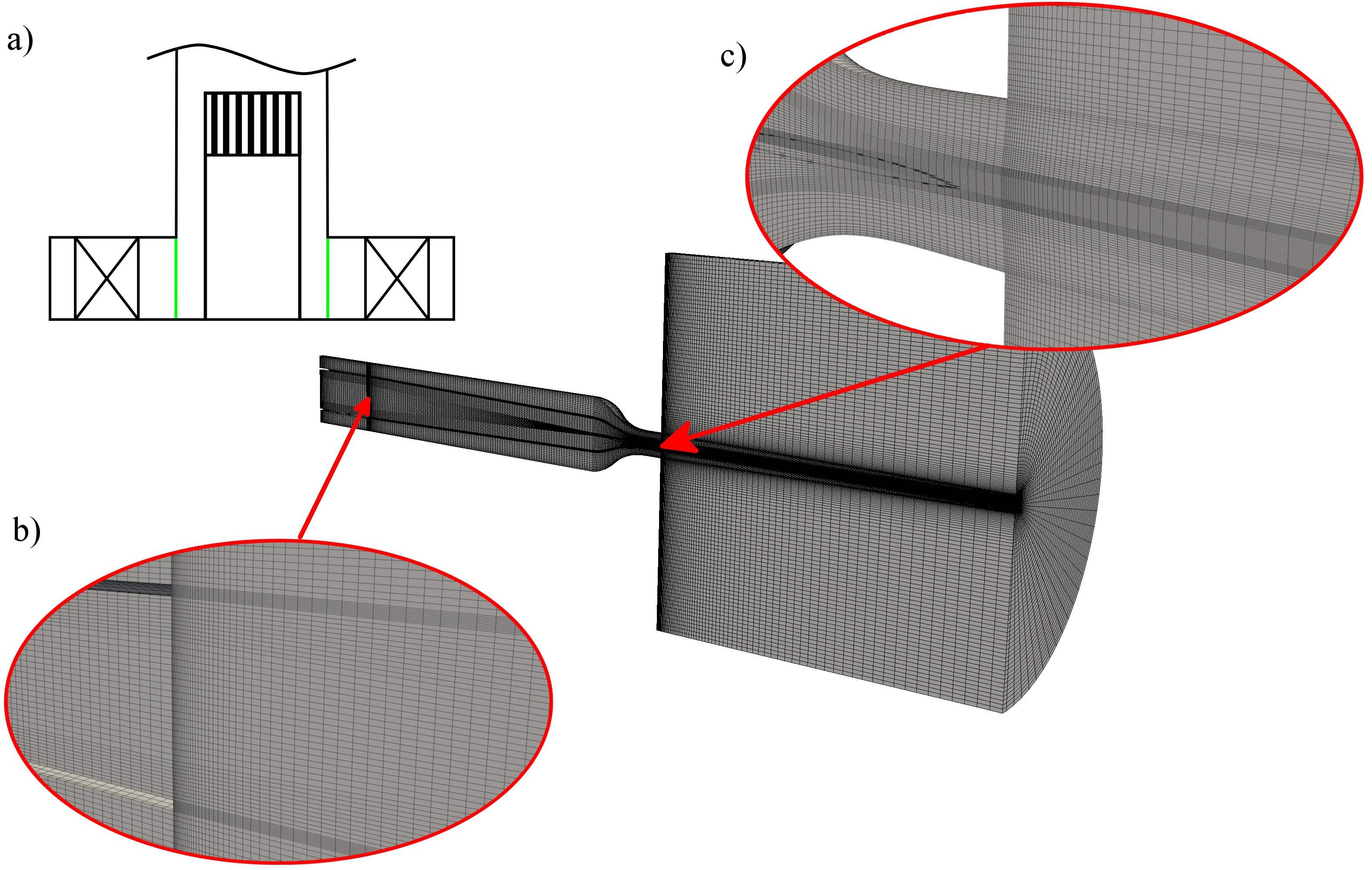}
\caption{Presentation of the mesh for the swirling jet.}
\label{fig:cfdGrid1} 
\end{figure}

The simulation domain was designed to emulate the experimental setup shown in \cref{fig:expFacility}. To ease the meshing procedure, the swirler with its vanes was omitted and the inflow was defined on a patch with the height of the swirler and the diameter of the following pipe. The green lines in \cref{fig:cfdGrid1} a) illustrate the position of this patch. \Cref{fig:cfdGrid1} b) gives a detailed view of the centerbody area. The honeycomb structure was not meshed explicitly, rather the volume occupied by the honeycomb in the experimental setup was modeled as an anisotropic porous medium. OpenFOAM allows for the inclusion of arbitrary body forces via the \textit{fvOptions} library. This was used to define a pressure loss in the $y$ and $z$ direction, without obstructing the flow in the $x$-direction. In the numerical simulations the unconfined domain had to be truncated at some radial and axial distance from the nozzle exit. A domain size of 15D in diameter and 15D in streamwise direction was found to be sufficient to render the simulation of the swirling jet in the near field of the nozzle insensitive to the boundary conditions imposed on the boundaries of the truncated domain. The mesh consisted in total of circa 1.2 million hexahedral cells. \\
The boundary conditions were defined in the following way: The velocities on the inlet boundary (\cref{fig:expFacility}, \cref{fig:cfdGrid1} a)) were obtained from a precursor simulation on a coarser tetrahedral mesh. This mesh included the complete swirler, with the swirl vanes set to the angle of the experiment. The bottom plate of the swirler, the inner and outer wall of the centerbody, the pipe, the nozzle and the end plate featured a no slip condition for the velocity. The unconfined domain was modeled with a mixed type boundary condition for the velocity. A zero gradient boundary condition was imposed on fluid leaving the computational domain, whereas fluid entering the domain was set to have a velocity of 0 m/s. The pressure was set to have a zero gradient across the inlet boundary, \textcolor{black}{normal to} all solid surfaces and was set to ambient pressure at the far field boundaries. The turbulence quantities were set with a fixed value boundary condition at the inlet patch and a zero gradient condition across the faces of the truncated unconfined domain. The behavior at solid faces was modeled with appropriate wall functions.

\subsection{Choice of turbulence model for URANS}
\label{sec:ransModel}
The prediction of a turbulent, highly swirling flow with such complex phenomena as vortex breakdown and the associated global mode is challenging for any turbulence model, because of the inhomogeneous turbulence structure and the inability of many turbulence models to account for rotation and curvature effects \cite{Shih.1997}. The results reported on the ability of different levels of turbulence modeling to predict these phenomena are not in unison. Guo et al. \cite{Guo.2001} investigated a swirling jet undergoing vortex breakdown in a sudden expansion using URANS with the standard \textit{k}-$\epsilon$ model. These authors achieve a very good prediction of the time-mean flow field, as well as the frequency of the global mode. Wegner et al. \cite{Wegner.2004} assess the performance of URANS to predict the flow in a movable block swirler device based on a comparison to LES and experiment. This study reports that simulations based on the standard \textit{k}-$\epsilon$ model showed the onset of the global mode in an early stage of their simulations, but as the simulation went on, this structure decayed and a steady state was reached. Moreover, Wegner et al. \cite{Wegner.2004} used a second-order Reynolds stress closure model. This model performed very well in predicting the time-mean flow and the frequency of the global mode. The work of \cite{Jochmann.2006} investigated the predictive performance of URANS based on the \textit{k}-$\epsilon$ model and a Reynolds stress model. Their benchmark were PIV and LDA measurements of a swirl-stabilized combustor. These authors report that the \textit{k}-$\epsilon$ model is able to capture the unsteady flow patterns they investigated very well. The difference between the two- and six-equation model is found to be a massive over-prediction of the eddy viscosity calculated by the \textit{k}-$\epsilon$ model.

The reasonable to good performance reported by \cite{Guo.2001} and \cite{Jochmann.2006} motivated us to use \textit{k}-$\epsilon$ models for the simulations in this work. However, the results produced by these simulations were in qualitative disagreement with the experiments. Neither the standard \cite{Jones.1972}, realizable \cite{Shih.1995} and the renormalization group \cite{Yakhot.1992} variant of the \textit{k}-$\epsilon$ model, nor the \textit{k}-$\omega$-SST \cite{Menter.1994} model and a nonlinear version of the \textit{k}-$\epsilon$ model \cite{Shih.1993} could reproduce vortex breakdown when used in computations involving the full domain (\cref{fig:cfdGrid1}). Most of these models were able to reproduce vortex breakdown, if the computational domain was reduced to the unconfined domain and measured time-mean velocity profiles were prescribed on the nozzle exit plane. However, the extent of the breakdown bubble, as well as the shear layer thicknesses were in significant disagreement with the measurements. In addition, no time-periodic global mode was evident from the simulation runs. The different performance in the different setups and the observation that the setup of Guo et al. \cite{Guo.2001} consisted of two pipes with an area jump and no nozzle, suggests that the anisotropic turbulence produced by the nozzle is one of the reasons for the poor performance of two-equation models in this study. Another contribution may come from the formation of a concentrated vortex core in the upstream part of the setup \cite{Rukes.2015}.

After two-equation models were found unsuitable for the simulation of the swirling jet, we attempted simulations with a Reynolds stress model following the observations of refs. \cite{Shih.1997,Wegner.2004,Jochmann.2006}. The Reynolds stress model of Launder et al. \cite{Launder.1975} was used for the simulations presented here. As discussed in \cref{sec:cfdVali}, a very convincing agreement between the simulated and measured time-mean and dynamical quantities was achieved with this model. 

\section{Data analysis}
\label{sec:dataAnalysis}
\subsection{Triple decomposition}
To study flows that feature a dominant coherent structure, it is convenient to adopt the triple decomposition of the flow field, introduced by \cite{Reynolds.1972}. The velocity vector is decomposed into a mean, a coherent, and a purely stochastic part,
\begin{equation}\label{eqn:triDeco}
\textbf{v}(\textbf{x},t) = \overline{\textbf{v}}(\textbf{x}) + \widetilde{\textbf{v}}(\textbf{x},t) + \textbf{v}'(\textbf{x},t).
\end{equation}
The time average is defined as
\begin{equation}\label{eqn:timeAv}
\overline{\textbf{v}}(\textbf{x}) = \lim\limits_{T \rightarrow \infty}{\frac{1}{T}\int_{0}^{T}{\textbf{v}(\textbf{x},t) \textrm{d}t}}.
\end{equation}
The coherent velocity component is obtained by subtracting the mean flow from the phase-averaged flow
\begin{equation}
 \widetilde{\textbf{v}}(\textbf{x},t) = \langle \textbf{v}(\textbf{x},t) \rangle - \overline{\textbf{v}}(\textbf{x}),
\end{equation}
with the definition of the phase average reading
\begin{equation}\label{eqn:phaseAv}
\langle \textbf{v}(\textbf{x},t) \rangle = \lim\limits_{N \rightarrow \infty}{\frac{1}{N}\sum_{0}^{N-1}{\textbf{v}(\textbf{x},t + n \tau)}},
\end{equation}
where $\tau$ denotes the period of the wave. In the present study, the experimental data set from which these quantities are to be extracted consists of non time-resolved uncorrelated PIV snapshots taken at arbitrary time increments. Therefore, no direct phase averaging is possible. However, Proper Orthogonal Decomposition allows for the \textit{a posteriori} reconstruction of the phase-averaged flow. This method is only applicable, if two coupled POD modes can be identified that span the subspace of the periodic structure from which the phase information can be retrieved. This has been demonstrated in the studies of \cite{Oberleithner.2011b, Stohr.2012}, where the POD was used to extract the dominant coherent structures from swirling jet experiments. Holmes et al. \cite{Holmes.1998b} can be consulted for a full account of the POD.

\subsection{Linear stability equations for turbulent flows}\label{sec:LSA}
The goal of this section is to derive the equations governing the stability of a coherent wave imposed on a time-mean flow. The starting point of the analysis are the incompressible Navier-Stokes and continuity equations
\begin{align}\label{eqn:nsStart}
\frac{\partial \textbf{v}}{\partial t} + (\textbf{v} \cdot \nabla)\textbf{v} &= -\nabla p +\frac{1}{\mathrm{Re}}\Delta \textbf{v}\\
\nabla \cdot \textbf{v} &= 0.
\end{align}
Classically, the linearized form of these equations is used to determine the growth of a disturbance on a laminar base flow that is a solution of the steady Navier-Stokes equations. This analysis breaks down when the disturbance grows to finite size and the nonlinear terms come into play.

To derive a system of equations that allows for the analysis of the fully saturated, turbulent state, the analysis has to rely on the time-mean, rather than the base flow \cite{Barkley.2006, Lugo.2014}. Similar to the Reynolds-averaging procedure, the triple decomposition, \cref{eqn:triDeco}, is substituted into \cref{eqn:nsStart}. The resulting system is then phase-averaged and thereafter time-averaged. The resulting system of equations reads \cite{Reynolds.1972}
\begin{align}
(\mv{v} \cdot \nabla) \mv{v} &= -\nabla \overline{p} -\frac{1}{\mathrm{Re}} \Delta \mv{v} - \nabla\cdot\underbrace{\overline{\fv{v}\fv{v}}}_{\tau^{R}} - \nabla\cdot\underbrace{\overline{\cv{v}\cv{v}}}_{\tau^{C}}\label{eqn:meanForced}\\
\nabla \cdot \mv{v} &= \nabla \cdot \cv{v} = \nabla \cdot\fv{v} = 0. 
\end{align}
Note that in addition to the usual Reynolds stresses $\tau^{R}$, stresses induced by the coherent fluctuation, $\tau^{C}$, are present. The time-mean flow is a solution of this steady forced equation, taking into account (nonlinear) modifications by the coherent and incoherent fluctuations. By subtracting \cref{eqn:meanForced} from the phase-averaged version of \cref{eqn:nsStart}, one arrives at the dynamical equation governing the evolution of an organized wave, growing on a nonlinearly corrected mean flow. This equation is given by
\begin{align}\label{eq:LSA}
\begin{split}
\frac{\partial \cv{v}}{\partial t} + (\mv{v}\cdot\nabla)\cv{v} + (\cv{v}\cdot\nabla)\mv{v} = -\nabla\widetilde{p} + \frac{1}{\mathrm{Re}}\Delta\cv{v} &+ \nabla\cdot\left(\overline{\cv{v}\cv{v}} - \cv{v}\cv{v} \right) \\
&- \nabla\cdot\underbrace{\left(\langle\fv{v}\fv{v}\rangle - \overline{\fv{v}\fv{v}} \right)}_{\widetilde{\tau}}.
\end{split}  
\end{align}
Focusing on a linear analysis, the quadratic term $\overline{\cv{v}\cv{v}} - \cv{v}\cv{v}$ is neglected. The term $\widetilde{\tau}$ represents the oscillation of the fine scale incoherent turbulence during the passage of the organized wave. From the energy considerations of Reynolds \& Hussain \cite{Reynolds.1972}, it is expected that $\widetilde{\tau}$ contributes at leading order and is hence retained in the analysis. Unfortunately, $\widetilde{\tau}$ is not known a priori and cannot be related to $\cv{v}$. Hence, specifying $\widetilde{\tau}$ poses a closure problem for the stability equations \eqref{eq:LSA}. 

Following Reau \& Tumin \cite{Reau.2002b}, closure is invoked by Boussinesq's turbulent viscosity hypothesis. The time-mean and phase-averaged fluctuations are specified via
\begin{equation}\label{eqn:phaseFluct}
\ \ \ \ \langle\textcolor{black}{\fv{v}\fv{v}}\rangle = \frac{2}{3}\langle k\rangle\textbf{I} - 2\nu_{t}^{p} \left(\nabla + \nabla^{\mathrm{T}}\right)\langle\textbf{v}\rangle
\end{equation}
and
\begin{equation}\label{eqn:timeFluct}
\overline{\fv{v}\fv{v}} = \frac{2}{3}\overline{k}\textbf{I} - 2\nu_{t}^{t}\left(\nabla + \nabla^{\mathrm{T}}\right)\mv{v},
\end{equation}
with $\nu_{t}$ denoting the eddy viscosity, $k$ the kinetic energy and \textbf{I} the identity matrix. The superscripts $t$, $p$ and T denote a time-averaged, phase-averaged and a transposed quantity, respectively. By time-averaging \cref{eqn:phaseFluct}, equating it to \cref{eqn:timeFluct} and by using the identity $\overline{\langle()\rangle} = \overline{()}$, Viola et al. show that $\nu_{t}^{t} = \nu_{t}^{p}$ \cite{Viola.2014}, and we can formally drop the superscripts. \Cref{eqn:phaseFluct} and \cref{eqn:timeFluct} yield the additional stress $\widetilde{\tau}$
\begin{equation}
\widetilde{\tau} = \frac{2}{3}\langle k\rangle\textbf{I} - \frac{2}{3}\overline{k}\textbf{I} - 2\nu_{t}\left(\nabla + \nabla^{\mathrm{T}}\right)\langle\textbf{v}\rangle + 2\nu_{t}\left(\nabla + \nabla^{\mathrm{T}}\right)\mv{v}. \label{eqn:tauStar2}
\end{equation}
We assume that phase and time-averaging changes the structure of the fine scale turbulence, but not the amplitude of the fluctuations. Therefore, $\langle k \rangle = \overline{k}$ and \cref{eqn:tauStar2} reduces to 
\begin{equation}
\widetilde{\tau} = 2\nu_{t}\left(\nabla + \nabla^{\mathrm{T}}\right)\cv{v}. \label{eqn:tauStar3}
\end{equation}
By introducing eq.~\eqref{eqn:tauStar3} into the governing dynamic equations \eqref{eq:LSA}, the eddy viscosity is lumped with the molecular viscosity into an effective viscosity $\nu_{\mathrm{eff}} = \nu + \nu_{t}$ and the final form of the stability equations reads
\begin{align}
 \frac{\partial \cv{v}}{\partial t} + (\mv{v}\cdot\nabla)\cv{v} + (\cv{v}\cdot\nabla)\mv{v} &= -\nabla\widetilde{p} + \nabla\cdot\left[\mathrm{Re}^{-1}_{\mathrm{eff}}\left(\nabla + \nabla^{\mathrm{T}}\right)\cv{v}\right]\label{eqn:nsFinal}\\
\nabla\cdot\cv{v} &= 0\label{eqn:contiFinal}.
\end{align}
Note that $\ts{Re}{eff}$ is now a function of space. 

\subsection{Solving the global mode with local LSA}
For local modal LSA, the ansatz for the disturbance quantities  
\begin{equation}
\left\{\widetilde{\textbf{v}}, \widetilde{p}\right\} = \left\{iF(r), G(r), H(r), P(r)\right\}\exp(i(\alpha x + m\theta -\omega t))
\end{equation}
is substituted into \cref{eqn:nsFinal} and \cref{eqn:contiFinal}. $\alpha$ and $m$ denote the axial and azimuthal wave number, $\omega$ signifies the frequency, indicating that the stability problem is solved in polar coordinates. The problem is completed by specifying appropriate boundary conditions. These are given by Khorrami et al. \cite{Khorrami.1989} as
\begin{equation}
F = G = H = P = 0
\end{equation}
for any azimuthal wave number $m$ at $r \rightarrow \infty$. Along the jet centerline the boundary conditions read, for $m = \pm 1$,
\begin{align}
\begin{split}
&F(0) \pm G(0) = 0\\
&H(0) = P(0) = 0.
\end{split}
\end{align}
The resulting eigenvalue problem constitutes a dispersion relation that combinations of $\alpha$ and $\omega$ have to satisfy for the shape functions $F, G, H$ and $P$ to be non-trivial solutions. Pseudospectral collocation techniques have proven to be well suited for the discretization of hydrodynamic stability problems \cite{Khorrami.1989}. The books of \cite{Boyd.2001,Trefethen.2000,Fornberg.1998} provide an excellent account of these methods. After discretization, the eigenproblems were solved with MATLABs \texttt{eig} and \texttt{eigs} function.

The eigenvalue problem can be solved for $\alpha$ given real or complex valued or for $\omega$ given \textcolor{black}{real}, resulting in a temporal, spatio-temporal or spatial stability analysis \cite{Huerre.1990}. The outcome of a spatio-temporal analysis is the absolute growth rate $\Im(\omega_{0})$ and frequency $\Re(\omega_{0})$. The parallel-flow profile is absolutely unstable, if the absolute growth rate is positive and convectively unstable, if the absolute growth rate is negative \cite{Huerre.1990}. The presence of a region of absolute instability is a necessary condition for the flow to sustain a global mode \cite{Batchelor.2003}.

The focus of this study lies on the ability of LSA to predict the frequency $\Re(\omega_{g})$ and growth rate $\Im(\omega_{g})$ of the global mode, as well as the location of frequency selection. These quantities are computed from a frequency selection criterion based on the absolute frequency $\omega_{0}$. In agreement with Pier \cite{Pier.2002}, we find that only the criterion of Chomaz et al. \cite{Chomaz.1991} applied to the time-mean flow provides predictions that are in agreement with experimental observations. Juniper et al. \cite{Juniper.2011} discuss practical aspects of the implementation of the frequency selection criterion.   

\subsection{Computing the turbulent viscosity from PIV data}
As outlined in Sec. \ref{sec:LSA}, Boussinesq's hypothesis is invoked to relate the coherent turbulent stresses $\widetilde{\tau}$ via the unperturbed eddy viscosity $\nu_{t}$ to the strain rate tensor of the mean field \cite{Reau.2002b}. This relationship is given by \cref{eqn:timeFluct} and reads in index notation   
\begin{align}
-\overline{v_{i}^{'}v_{j}^{'}} + \frac{2}{3}\overline{v_{k}^{'}v_{k}^{'}}\delta_{ij}&=  \nu_{t}\left[\frac{\partial \overline{v}_{i}}{\partial x_{j}} + \frac{\partial \overline{v}_{j}}{\partial x_{i}}\right].\label{eqn:eddy3d}
\end{align}
This expression cannot be used for a direct computation of the eddy viscosity. In a swirling flow undergoing vortex breakdown, the normal and shear stresses are of the same order of magnitude and no simplifications of the Reynolds stress tensor are generally valid. \Cref{eqn:eddy3d} results in an overdetermined system of equations for $\nu_{t}$ that can only be solved in an approximate least-squares sense \cite{Ivanova.2012}, reading
\begin{equation}\label{eqn:LSQ}
\nu_{t} = \frac{\left(-\overline{v_{i}^{\prime}v_{j}^{\prime}}+\frac{2}{3}\overline{v_{k}^{'}v_{k}^{'}}\delta_{ij}\right)\cdot\left(\frac{\partial \overline{v}_{j}}{\partial x_{i}}+\frac{\partial \overline{v}_{i}}{\partial x_{j}}\right)}{\left(\frac{\partial \overline{v}_{l}}{\partial x_{m}} + \frac{\partial \overline{v}_{m}}{\partial x_{l}}\right) \cdot \left(\frac{\partial \overline{v}_{l}}{\partial x_{m}} + \frac{\partial \overline{v}_{m}}{\partial x_{l}}\right)}.
\end{equation}

One concern with this approach is that the solution of the overdetermined system of equations \cref{eqn:LSQ} is not bounded. It is observed in this study that the solution of \cref{eqn:LSQ} yields regions of negative eddy viscosity. The problem is worsened by the fact that the computed eddy viscosity varies rapidly in the PIV domain, producing steep gradients. This is problematic from the point of view of a numerical implementation, because the rapid change in viscosity can lead to convergence problems with the eigenvalue solver. Conceptually, a negative eddy viscosity is also questionable. In fact, steps are taken in RANS modeling to ensure that the modeled turbulence quantities are strictly positive \cite{Allmaras.2012,Ilinca.1996}, otherwise the computation may suffer an immediate and irrecoverable breakdown \cite{Ilinca.1996}. In this study a clipping strategy was attempted that resets negative values of the eddy viscosity to zero. It should be noted that a negative eddy viscosity may be well embedded, if an explicit separation of scales in the turbulent spectrum is introduced. This is classically done in Large Eddy Simulations. There, a negative eddy viscosity can be interpreted as backscatter of turbulent kinetic energy from small turbulent scales to energy containing, larger scales \cite{Kraichnan.1976,Shah.1995}. However, even in the context of Large Eddy Simulations, clipping or averaging of negative eddy viscosity is common practice to avoid numerical instabilities \cite{Rodi.1997}.

In order to avoid the problem of negative eddy viscosities, another approach was tested for the experimental data. Returning to the ideas of turbulence modeling in CFD, the eddy viscosity is computed in the $k-\epsilon$ model family via
\begin{equation}\label{eqn:ke}
\nu_{t} = C_{\mu}\frac{k^{2}}{\epsilon}.
\end{equation}
$\epsilon$ denotes the turbulent dissipation rate and $C_{\mu}$ is a model constant. We assume the value of 0.09 for $C_{\mu}$, which is the value of the baseline $k-\epsilon$ model \cite{Jones.1972}. Both, $k$ and $\epsilon$ can be computed from PIV data and are strictly positive. The necessity for clipping is thus avoided.

A method for computing $\epsilon$ from PIV data was proposed by Sheng et al. \cite{Sheng.2000}. This approach is called a PIV-LES method by these authors. They observe that Large Eddy Simulations (LES) and PIV are similar in that both methods have a finite spatial resolution that is usually larger than the smallest scales in the flow. The dissipation of turbulent eddies, however, takes place at the smallest scales. Sheng et al. suggest to adopt the subgrid scale modeling of LES to PIV, in order to improve the turbulent dissipation rate estimation from measurements. The dissipation rate is estimated according to
\begin{equation}
\epsilon \approx \epsilon^{\mathrm{SUB}} = -2\overline{\tau_{ij}^{\mathrm{SUB}} S^{\mathrm{GR}}_{ij}}.
\end{equation}
The superscript SUB refers to a \textbf{SUB}grid scale quantity that needs to be modeled and the superscript GR refers to a \textbf{G}rid \textbf{R}esolved quantity. $S$ denotes the strain rate tensor. For brevity of notation $\epsilon^{\mathrm{SUB}}$ is denoted by $\epsilon$ in the following. Sheng et al. discuss different options to model the subgrid scale stresses $\tau_{ij}^{\mathrm{SUB}}$. We found no significant difference between the different subgrid scale models discussed by Sheng et al. and therefore stick to the widely used model proposed by Smagorinsky \cite{Smagorinsky.1963}. The dissipation rate is then computed via
\begin{equation}
\epsilon = 2\overline{C_{s}^2\delta^{2}_{x}\left|\textbf{S}^{\mathrm{GR}}\right|S_{ij}^{\mathrm{GR}}S_{ij}^{\mathrm{GR}}}.
\end{equation}  
$C_{s}$ is a model constant with the value 0.17 and $\delta_{x}$ is a measure of the grid size. $|...|$ refers to the absolute value. 

\section{Results - Experimental investigation}
\label{sec:results}
\subsection{The time-mean flow}
\begin{figure}
\centering
\includegraphics{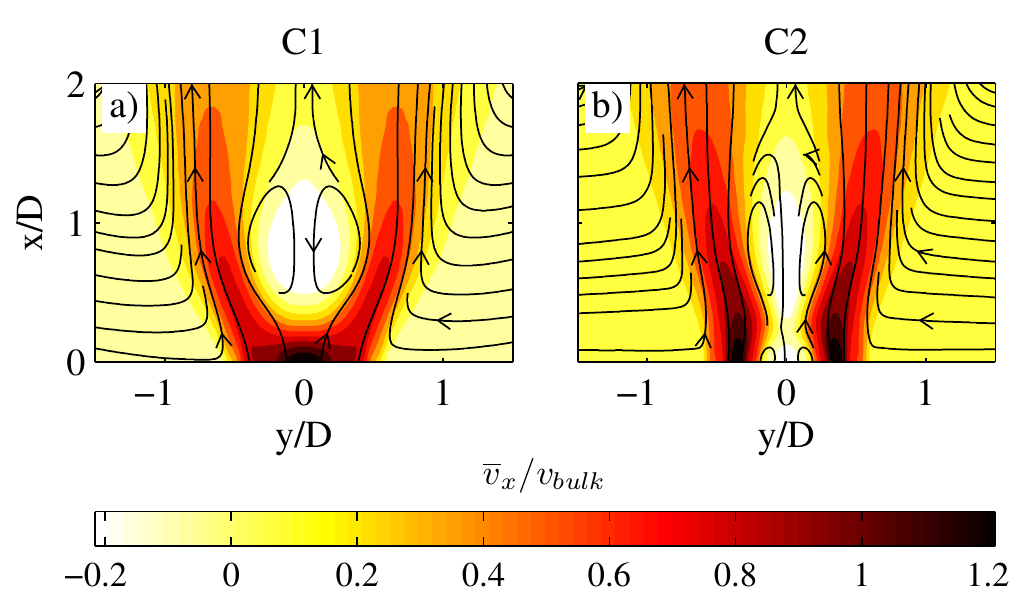}
\caption{The time-mean axial velocity fields of C1 (a)) and C2 (b)) as measured with PIV. All velocities are normalized with their respective bulk velocity.}
\label{fig:meanVelo} 
\end{figure}

This section introduces the time-averaged velocity fields of C1 and C2 as measured with PIV. As shown in \cref{fig:meanVelo} a), C1 exhibits the typical velocity field of a swirling jet undergoing vortex breakdown. The incoming jet is forced to an expansion around the recirculation bubble. C2 features a time-mean velocity field that is often found in swirl stabilized combustion research (\cref{fig:meanVelo} b)), where centerbodies are used for flame stabilization or pilot fuel injection. The velocity field is more parallel and the internal recirculation zone features the wake of the centerbody and the recirculation bubble of vortex breakdown.     

\subsection{Dynamic properties of the global mode}
\begin{figure}
\centering
\includegraphics{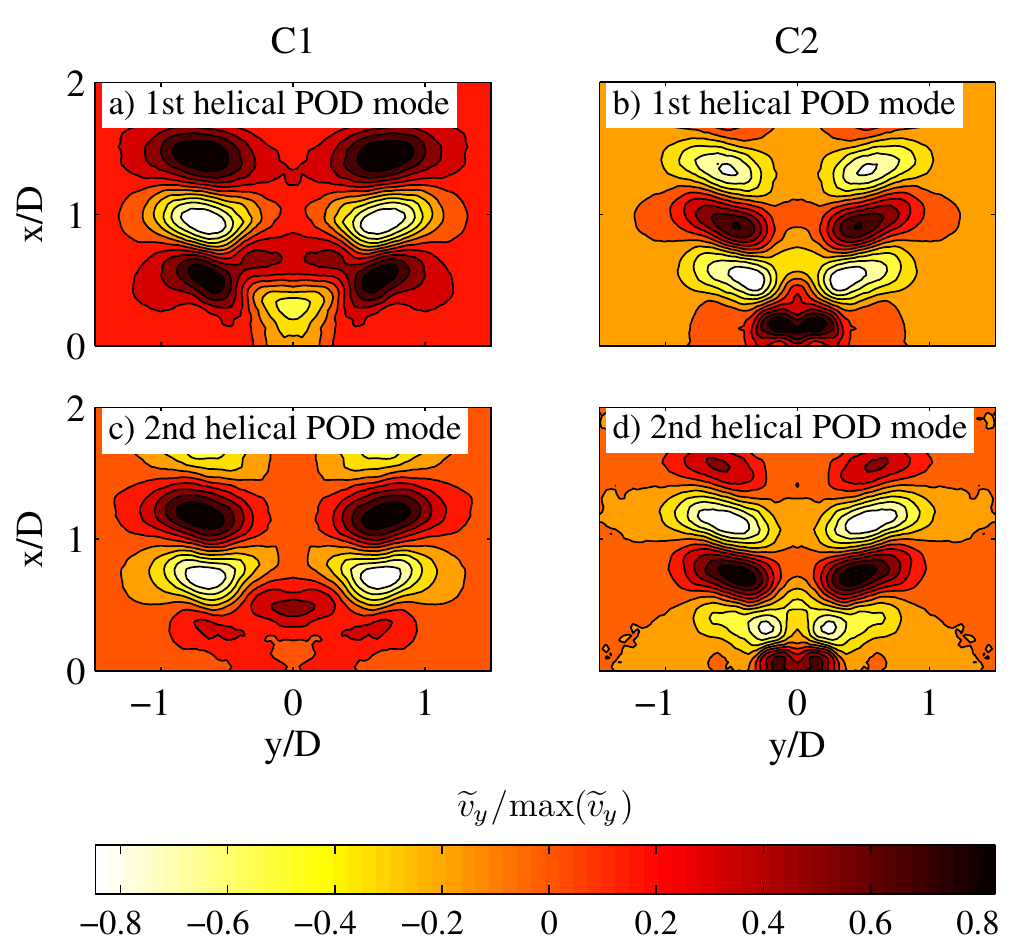}
\caption{The cross stream velocity component of the dominant POD modes of C1 (a), c)) and C2 (b), d)) derived from PIV measurements.}
\label{fig:coheVelo} 
\end{figure}

\begin{figure}
\centering
\includegraphics{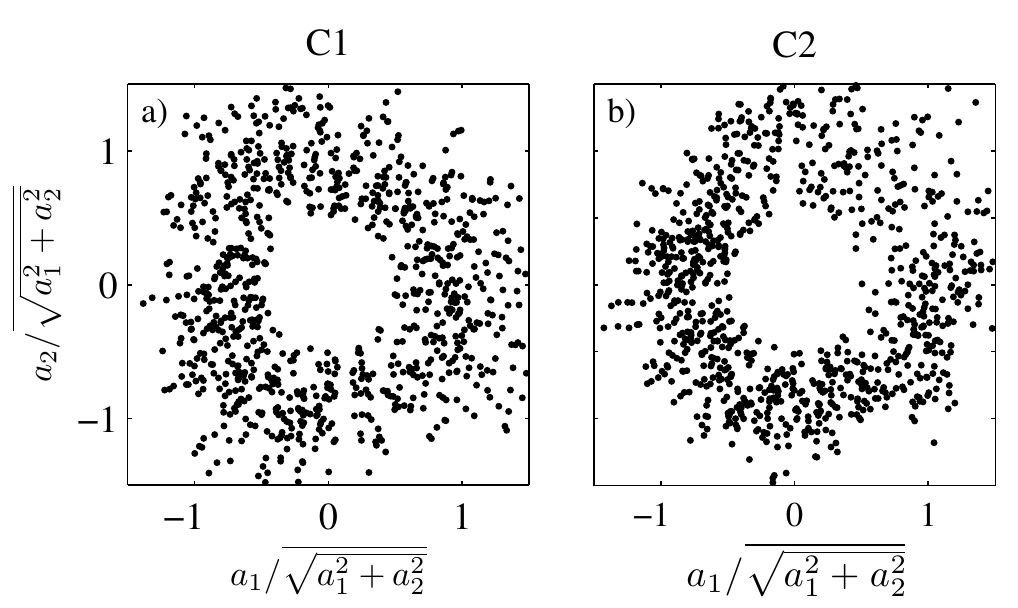}
\caption{Phase portrait of the POD modal amplitudes of C1 (a)) and C2 (b)) derived from PIV measurements.}
\label{fig:phasePort} 
\end{figure}

A mode pair that spans the subspace of the periodic structure was identified from a POD analysis of the measurements of C1 and C2. This is evident from the coherent $\widetilde{v}_{y}$ component of the respective mode (\cref{fig:coheVelo} a), c) for C1 and b), d) for C2) and the circular phase portrait of the modal coefficients (\cref{fig:phasePort}) \cite{Oberleithner.2011b}. These results confirm that the dynamics of each configuration are governed by the same periodic structure.
Differences between the structures of C1 and C2 prevail in the axial wavelength which is 10\% smaller for the modes of C2 than those of C1. The difference in axial wavelength is related to a different temporal dynamic of the structure. From the time resolved measurements of C1 and C2 (LDA and pressure), the frequency of the global mode is calculated as 47 Hz for C1 and 76 Hz for C2.  

\subsection{The eddy viscosity fields}
\begin{figure}
\centering
\includegraphics{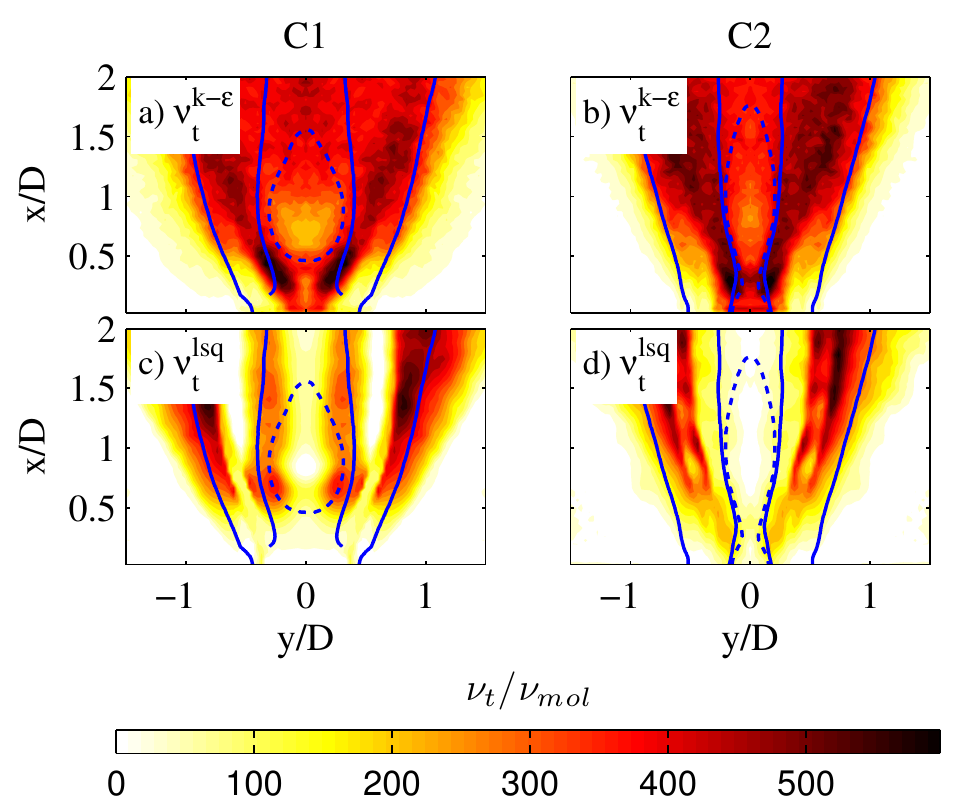}
\caption{The eddy viscosity derived from the $k-\epsilon$ model approach and from the least squares fit to all resolved Reynolds stress components of the PIV measurements. a) and c) present the $k-\epsilon$ and least-squares approach for C1 and b) and d) for C2. The position of the inner and outer shear layer is given for reference by the solid line. The dotted lines indicate the contour line of zero axial velocity.}
\label{fig:eddyField} 
\end{figure}

The results of the $k-\epsilon$ approach and the least-square fit are discussed in this section. The respective eddy viscosities are denoted by $\nu_{t}^{k-\epsilon}$ and $\nu_{t}^{lsq}$. The $\nu_{t}^{k-\epsilon}$ for C1 derived from \cref{eqn:ke} is presented in \cref{fig:eddyField} a) and $\nu_{t}^{lsq}$ computed according to \cref{eqn:LSQ} is shown in c). The contribution of the global mode was removed from the fluctuating velocity fields by computing a POD and reconstructing the fluctuating velocity with the corresponding POD modes excluded. $\nu_{t}^{k-\epsilon}$ is distributed almost homogeneously for $x/\mathrm{D} > 0.5$ within the jet. Note that $\nu_{t}^{k-\epsilon}$ takes on large values upstream of the recirculation bubble along the inner shear layers. $\nu_{t}^{lsq}$ has a significantly different spatial distribution (\cref{fig:eddyField} c)). $\nu_{t}^{lsq}$ deviates strongly from the molecular viscosity for $x/\mathrm{D} > 0.5$, but remains relatively small further upstream. Particularly, $\nu_{t}^{lsq}$ differs significantly along the inner shear layer and takes on its largest values in the vicinity of the outer shear layer. In the annular jet, for $x/D > 0.5$, and in the wake of the breakdown bubble, only small values of $\nu_{t}^{lsq}$ are evident.    
$\nu_{t}^{k-\epsilon}$ of C2 in \cref{fig:eddyField} b) is distributed similarly to C1 (\cref{fig:eddyField} a)). $\nu_{t}^{k-\epsilon}$ has large values in the bulk of the flow, in the inner and outer shear layers and particularly upstream of the recirculation bubble. $\nu_{t}^{lsq}$ in \cref{fig:eddyField} d) is less aligned with the shear layer position compared to C1. As in \cref{fig:eddyField} c), the largest values are attained for $x/\mathrm{D} > 0.5$.

\subsection{Regions of absolute instability}
\begin{figure}
\centering
\includegraphics{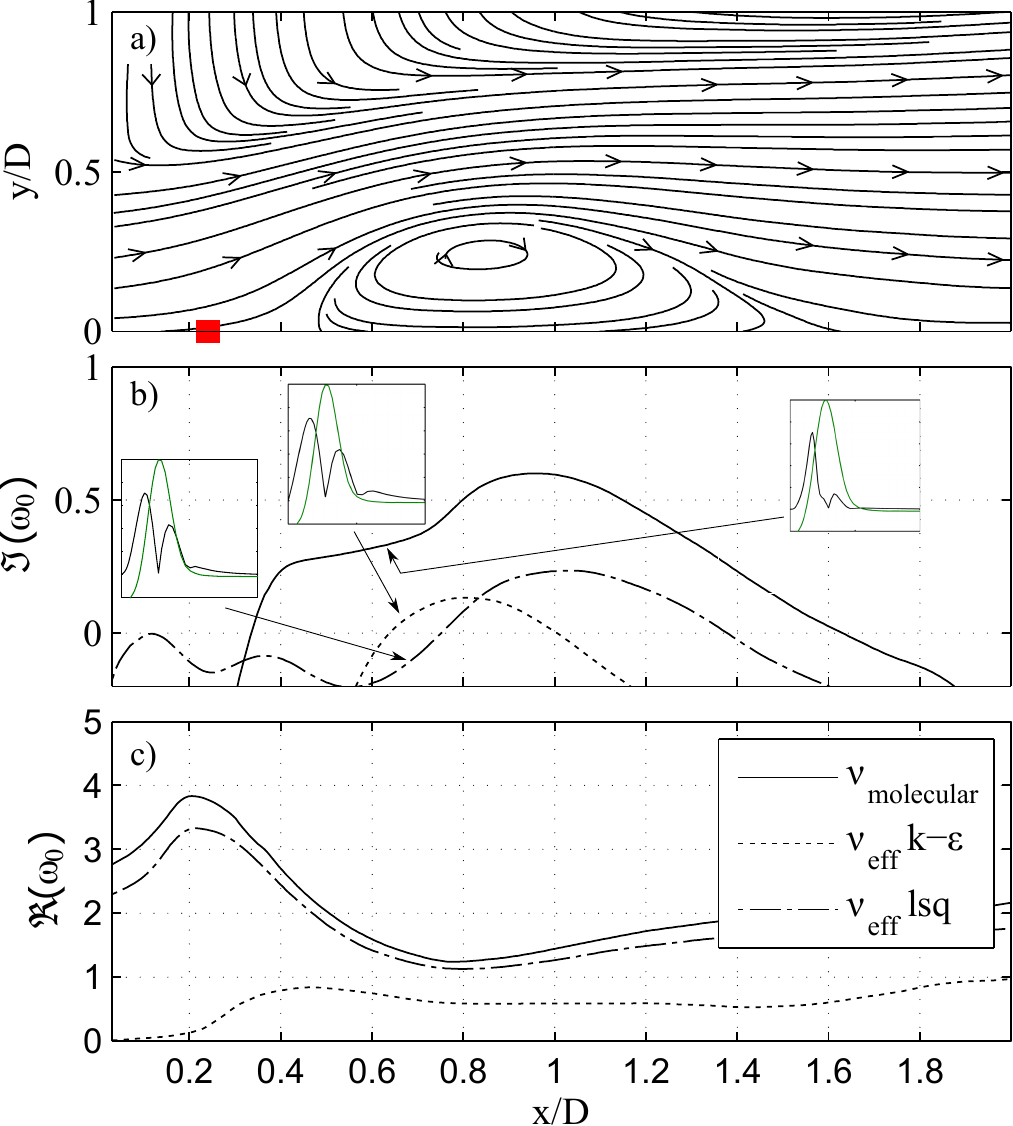}
\caption{LSA results for C1 (PIV). a): Streamlines indicate the time-mean flow. The red square marks the axial position of the wavemaker \textcolor{black}{of the $\nu_{eff}^{lsq}$ computation}. b) Absolute growth rate curves computed for different viscosities. \textcolor{black}{The insets show the absolute value of the eigenfunction $H(r)$ (black) and the axial velocity profile (green) at x/D = 0.7.}  c) The real part of the absolute \textcolor{black}{frequency} for the different viscosities. \textcolor{black}{The legend also applies to b).}}
\label{fig:sTempJet} 
\end{figure}

The impact of the different effective viscosities on the predictions of the LSA is outlined here. For reference, each analysis is supplemented with the results of LSA based only on the molecular viscosity $\nu_{mol}$. The results of LSA applied to the measured time-mean flows of C1 are presented in \cref{fig:sTempJet}. The \textcolor{black}{streamlines of the time mean velocity field of} C1 is given for reference in \cref{fig:sTempJet} a). The impact of the eddy viscosity on the region of absolute instability is presented in \cref{fig:sTempJet} b). Switching from the molecular viscosity to the effective viscosity has a significant impact on the absolute growth rate curve. The inclusion of $\nu_{t}^{k-\epsilon}$, as well as $\nu_{t}^{lsq}$, substantially reduces the region of absolute instability in the flow. Note that the absolute growth rate \textcolor{black}{that results from the inclusion of $\nu_{t}^{lsq}$} is brought close to zero for $x/\mathrm{D} < 0.3$ compared to the computation based on $\nu_{mol}$. In addition, the absolute frequency strongly depends on the way the eddy viscosity is computed (\cref{fig:sTempJet} c)). $\nu_{t}^{lsq}$ leads to a distribution of the absolute frequency that resembles the computation with $\nu_{mol}$ very closely, with a small reduction in the largest value of the frequency. Applying $\nu_{t}^{k-\epsilon}$ to the LSA computation leads to a profound change in the absolute frequency curve. The peak in the curve at around $0.2\mathrm{D}$ is completely eliminated and values of the absolute frequency downstream of $0.2\mathrm{D}$ are strongly reduced compared to the other two computations.\\
\textcolor{black}{This change in frequency is peculiar and one may wonder whether the instability is still generated by the same mechanism and flow region. The insets in \cref{fig:sTempJet} b) show that all eigenfunctions peak in the inner shear layer and the presence of one viscosity or another does not substantially change the mode shape. Hence, the choice of the eddy viscosity significantly affect the growth rate {\itshape and} frequency of the stability mode.}

Next, C2 is considered. In addition to the eddy viscosities presented for C1 in \cref{fig:sTempJet}, an additional computation of the $k-\epsilon$ based approach is included. \textcolor{black}{This is motivated by the wrong prediction of the global modes growth rate and frequency based on $\nu_{t}^{k-\epsilon}$ observed for C1. These are discussed in detail in the subsequent chapter.} A source of error in the estimation of the turbulent dissipation rate from PIV data is the spatial resolution of the measurement. Delafosse et al. \cite{Delafosse.2011} compare the turbulent dissipation rate estimations from PIV measurements at different spatial resolutions and report an increase by 220\% of the turbulent dissipation rate, if the spatial resolution is \textcolor{black}{improved} by a factor of two. Tanaka \& Eaton \cite{Tanaka.2007} assess the accuracy of turbulent dissipation rate predictions obtained from a PIV evaluation of synthetic vector fields. They find an optimal PIV resolution in the range of $\eta/10$ to $\eta/2$, where $\eta$ is the Kolmogorov scale. If this resolution is exceeded, the dissipation rate is underestimated. The spatial resolution in this study is constrained by the need to include the spatial evolution of the entire global mode and is, thus, larger than the suggested optimum range. However, the spatial resolution of the PIV evaluation does not only depend on the size of the region of interest and the camera resolution, but also on the interrogation window size of the PIV evaluation. Following the recommendations of Saarenrinne \& Piirto \cite{Saarenrinne.2000}, we re-evaluated the PIV data with an interrogation window size of 16x16 pixel, in addition to the 32x32 pixel window size.

\begin{figure}
\centering
\includegraphics{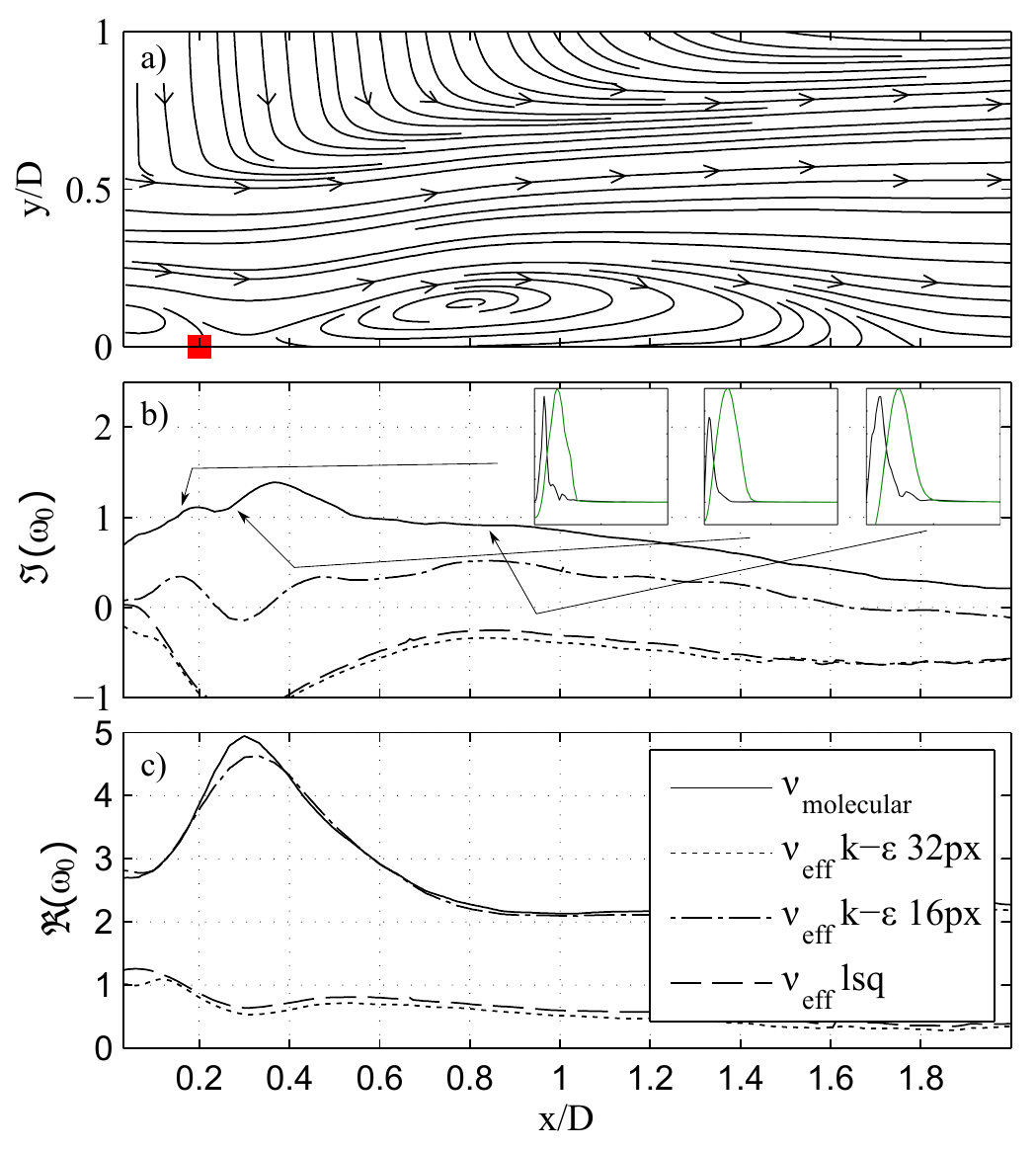}
\caption{LSA results for C2 (PIV). a): Streamlines indicate the time-mean flow. The red square marks the axial position of the wavemaker \textcolor{black}{of the $\nu_{eff}^{lsq}$ computation}. b) Absolute growth rate curves computed for different viscosities. \textcolor{black}{The insets show the absolute value of the eigenfunction $H(r)$ (black) and the axial velocity profile (green) at x/D = \{0.17, 0.33, 0.84\}.} c) The real part of the absolute \textcolor{black}{frequency} for the different viscosities. \textcolor{black}{The legend also applies to b).}}
\label{fig:sTempWake} 
\end{figure}

The impact of the different eddy viscosities on the absolute growth rate curves of C2 is shown in \cref{fig:sTempWake} b). The \textcolor{black}{streamlines of the} time-mean velocity field is given for reference in \cref{fig:sTempWake} a), where the wavemaker is marked by the red square. We find that as for C1 the extent of the region of absolute instability is reduced by the inclusion of $\nu_{t}^{lsq}$. $\nu_{t}^{k-\epsilon}$ completely suppresses the absolute instability for both resolutions considered. The absolute frequency is only mildly influenced by $\nu_{t}^{lsq}$, but strongly reduced for both $k-\epsilon$ computations (\cref{fig:sTempWake} c)). \\
\textcolor{black}{It is interesting to note that the computations based on the molecular viscosity show the largest absolute growth rate at $x/D \approx 0.4$. This is located between the vortex breakdown bubble and the wake of the centerbody, where flow recirculation is relatively weak. This raises the question whether the absolute instability is driven by the same mode throughout the flow. \Cref{fig:sTempWake} b) shows disturbance eigenfunctions at several axial stations. Note that the eigenvectors are normalized arbitrarily and \cref{fig:sTempWake} b) does not warrant the conclusion that the instability is at one axial location ''stronger`` than at the other. Clearly, at every streamwise station, the unstable mode is related to the inner shear layer. }

 
\subsection{Frequency and growth rate predictions}
The prediction of the global mode growth rate and shedding frequency are discussed next. \Cref{tab:freqSelMeas} shows the predictions of the global mode for C1 and C2, with $\mathrm{St} = f\mathrm{D}/v_{bulk}$ and $f = \Re(\omega_{g})/2\pi$. The predictions are compared to the Strouhal number derived from time resolved measurements for C1 and C2.

Considering C1 first, the prediction of the LSA based on $\nu_{mol}$ is in reasonable agreement with the measured value at a deviation of 19\%. The frequency prediction is significantly improved by including $\nu_{t}^{lsq}$. In this case, the predicted frequency deviated by 7\% from the measured value. The LSA calculation with $\nu_{t}^{k-\epsilon}$ severely under-predicts the measured frequency, with a deviation of 88\%. Similar observations can be made for C2. The calculations based on $\nu_{mol}$ and $\nu_{t}^{lsq}$ agree to the measured value within 7\% and 3.5\%. The prediction based on $\nu_{t}^{k-\epsilon}$ is off by 84\% for an interrogation window of 32x32 pixel. Given that the frequency selection follows from the absolute growth rate curves and given the similarity of the growth rate curves for the 32x32 pixel and the 16x16 pixel case, we have forgone the calculation of the frequency selection for the 16x16 pixel case.

The predicted growth rates of the global mode are indicated by $\Im(\omega_{g})$ in \cref{tab:freqSelMeas}. They are compared to the zero growth rate of the global mode on the limit cycle. \textcolor{black}{The notion that the flow is marginally stable with a growth rate of zero is supported by the several theoretical analyses \cite{Noack.2003,Lugo.2014,Barkley.2006,Turton.2015}}. For both, C1 and C2, the $k-\epsilon$ based calculations predict a negative growth rate that is significantly below zero. The computations based on $\nu_{mol}$ also predict a growth rate that is substantially smaller than zero. For both cases, the computations with $\nu_{t}^{lsq}$ yield growth rates that are close to zero.\\ It is astonishing that LSA predictions \textcolor{black}{of the frequency and growth rate} based on a strongly non-parallel and highly turbulent three-dimensional flow are in that good agreement with experimental observations \textcolor{black}{and theoretical predictions}, indicating the physical relevance of time-mean flow stability analysis.
\begin{table}
\centering
\begin{tabular}{c c c c c c}
\toprule
                      &                     &                         & St         & $\Im(\omega_{g})$ &  $\ts{x}{WM}/\mathrm{D}$\\
\midrule
\multirow{3}{*}{C1}   &  \ldelim\{{3}{13pt} & $\nu_{mol}$       & 0.49 (0.41) & -0.13 (0)         & 0.2 \\
                      &                     & $\nu_{eff}^{k-\epsilon}$ & 0.05 (0.41) & -0.23 (0)         & 0.25 \\
                      &                     & $\nu_{eff}^{lsq}$        & 0.44 (0.41) & -0.01 (0)         & 0.24 \\
\hline
\multirow{3}{*}{C2}   &  \ldelim\{{3}{13pt} & $\nu_{mol}$              & 0.55 (0.51) & -0.27 (0)         & 0.25 \\
                      &                     & $\nu_{eff}^{k-\epsilon}$ & 0.08 (0.51) & -0.18 (0)         & 0.3 \\
                      &                     & $\nu_{eff}^{lsq}$        & 0.52 (0.51) & -0.07 (0)         & 0.2 \\
\bottomrule
\end{tabular}
\caption{The Strouhal number predictions of the global mode of C1 and C2 compared to the measured value which is given in brackets. The predicted growth rate $\Im(\omega_{g})$ is compared to the theoretical on the limit cycle of the global mode (given in brackets). $\omega_{g}$ is normalized via $\omega_{g}\mathrm{D}/v_{bulk}$. The wavemaker location is indicated by $\ts{x}{WM}$. All shown numbers correspond to PIV measurements. \label{tab:freqSelMeas}}   
\end{table}

\begin{figure}
\centering
\includegraphics{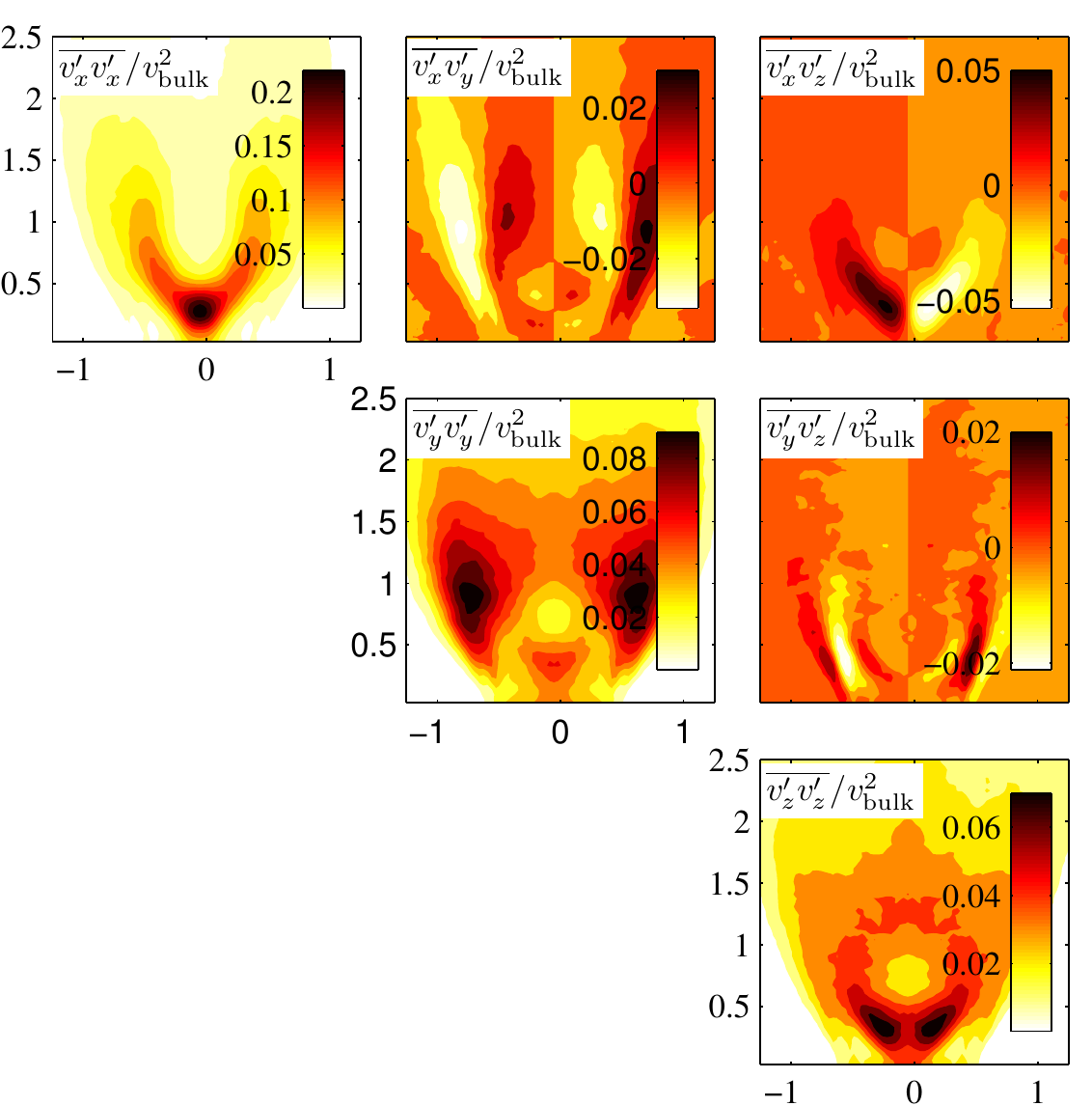}
\caption{The components of the Reynolds stress tensor in the PIV domain for C1.}
\label{fig:reStressC1} 
\end{figure}

We conclude this section by returning to the predictive quality of LSA calculations based on $\nu_{t}^{k-\epsilon}$. We argue that there is an intrinsic deficiency to the predictions based on the $k-\epsilon$ model. The intrinsic problem is the focus on the normal stresses that contribute to $k$. In swirling jets undergoing vortex breakdown, the off-diagonal elements of the Reynolds stress tensor are of a similar magnitude as the main-diagonal elements, as is illustrated for C1 in \cref{fig:reStressC1}. The normal stresses are particularly large upstream of the recirculation bubble (cf. \cref{fig:eddyField} a) and \cref{fig:reStressC1}). This is related to the unsteady movement of the recirculation bubble (\cite{Rukes.2015} can be consulted for details). From \cref{fig:eddyField} a) it is evident that the eddy viscosity takes on large values in the inner shear layer. We have identified in another study \cite{Rukes.2015b} that the growth of unstable modes in the inner shear layer relates to the formation of the global mode. \Cref{fig:sTempJet} and \cref{fig:sTempWake} a) show that the wavemaker is located upstream of the recirculation bubble and in regions, where a very large value of $\nu_{t}^{k-\epsilon}$ is present. The large value of $\nu_{t}^{k-\epsilon}$ suppresses the growth of the unstable modes in the LSA, which in turn determine the wavemaker location and the value of the frequency and growth rate at the wavemaker location, and thus deteriorates the predictive capability. 

\section{Results - Experiment vs URANS}
\label{sec:results2}
Is the turbulence model identified for the analysis of swirling jets well suited for the prediction of such flows in URANS? This question poses a consistency check for the line of thought presented so far. If the full Reynolds stress tensor needs to be taken into account for an accurate \textit{a posteriori} analysis of the flow's dynamic features, it should also be taken into account in the prediction of the flow.

Incidentally, URANS provides data upstream the region of interest of PIV measurements. This is very valuable if a global stability analysis is considered that requires upstream and downstream boundary conditions that are placed sufficiently far away from the domain of the global mode.  
\subsection{The time-mean flow}
\begin{figure}
\centering
\includegraphics{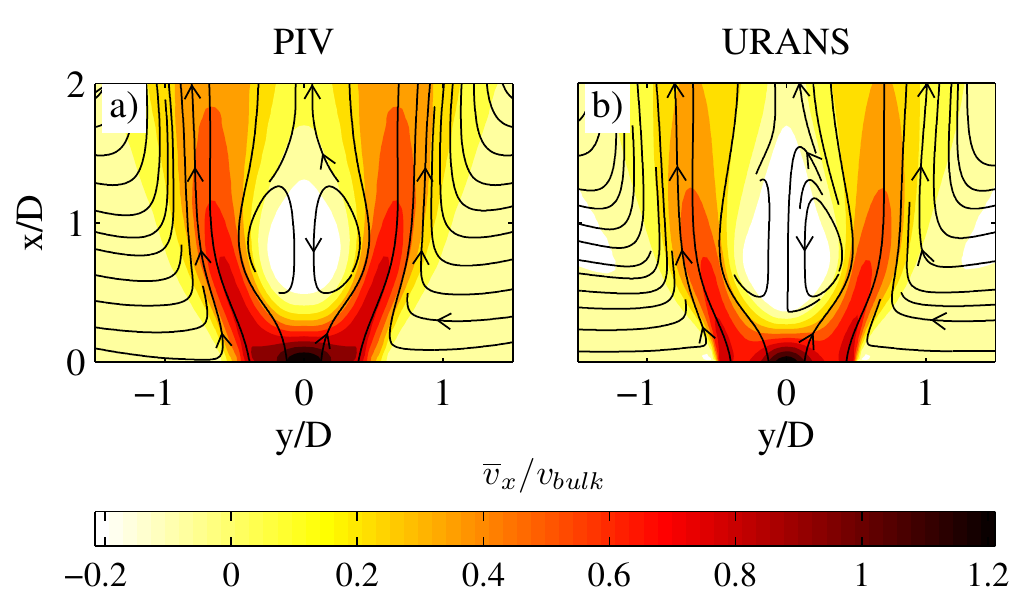}
\caption{The time-mean axial velocity fields of C1 from PIV (a)) and URANS (b)). All velocities are normalized with their respective bulk velocity.}
\label{fig:meanVelo2} 
\end{figure}

This section briefly introduces the time-mean axial velocity fields of C1 computed from URANS and compared to PIV. A detailed comparison of the URANS and PIV results is provided in \cref{sec:cfdVali}. The axial velocity fields of \cref{fig:meanVelo2} a) and b) show that the URANS computations are in good agreement with the PIV measurements and further analysis can be reliably conducted based on the URANS results.          

\subsection{Regions of absolute instability}
\begin{figure}
\centering
\includegraphics{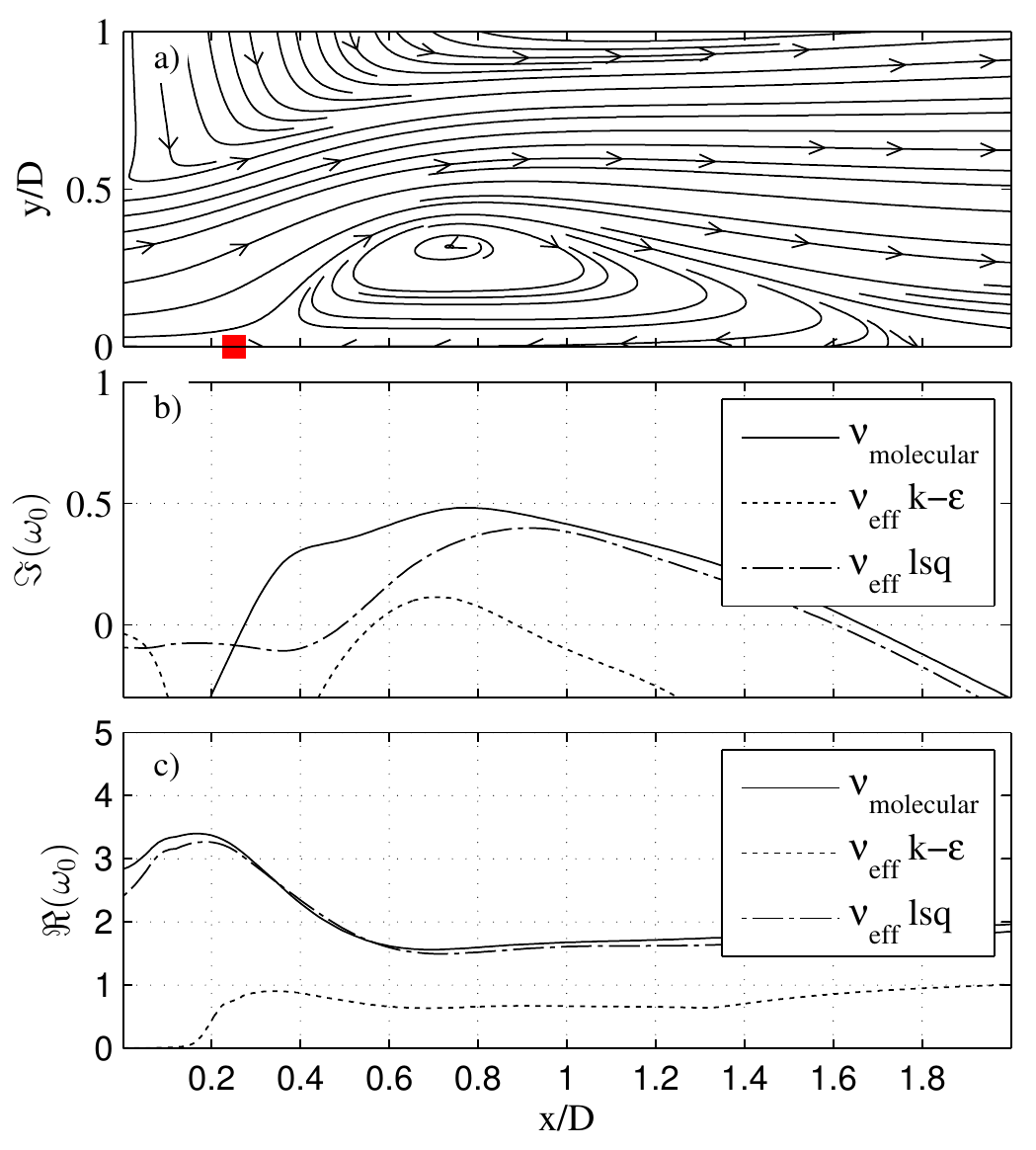}
\caption{LSA results for C1 (URANS). a): Streamlines indicate the time-mean flow. The red square marks the axial position of the wavemaker \textcolor{black}{of the $\nu_{eff}^{lsq}$ computation}. b) Absolute growth rate curves computed for different viscosities. c) The real part of the absolute \textcolor{black}{frequency} for the different viscosities.}
\label{fig:sTempURANS} 
\end{figure}

\Cref{fig:sTempURANS} a) shows the \textcolor{black}{streamlines of the time-mean velocity field together} with the wavemaker location that is marked by the red square. Note that the values of $k$ and $\epsilon$, as well as all components of the Reynolds stress tensor are directly obtained from the simulation. \Cref{eqn:ke} and \cref{eqn:LSQ} are then applied to these quantities to derive the eddy viscosity. We find a behavior that is similar to the LSA of the experimental data. The region of absolute instability is reduced by the inclusion of the eddy viscosity. This effect is particularly strong with $\nu_{t}^{k-\epsilon}$. The modification of the absolute growth rate curves is indicated in \cref{fig:sTempURANS} c). The use of $\nu_{t}^{lsq}$ reduces the absolute frequency to a small extent, whereas $\nu_{t}^{k-\epsilon}$ strongly suppresses the absolute frequency.\\ \textcolor{black}{Similar to C1 (PIV), we find by inspection of the eigenfunctions (not shown) that the absolute instability is related to the inner shear layer.}

\subsection{Frequency and growth rate predictions}
We turn to the global mode growth rate and frequency predictions of the LSA. The values obtained for the PIV measurements are given for reference in \cref{tab:freqSelCFD}. Beginning with the Strouhal number predictions, the LSA based on $\nu_{mol}$ results in a deviation of 28\% from the measured value. The Strouhal number prediction strongly improves when the eddy viscosity is based on the least squares fit. In this case, the predicted and measured frequency agree within 7\%. The prediction from LSA based on the $k-\epsilon$ computations is off by 85\%.
\begin{table}
\centering 
\begin{tabular}{c c c c c c}
\toprule
                            &                     &                         & St         & $\Im(\omega_{g})$ &  $\ts{x}{WM}/\mathrm{D}$\\
\midrule
\multirow{3}{*}{C1 PIV}     &  \ldelim\{{3}{11pt} & $\nu_{mol}$       & 0.49 (0.41) & -0.13 (0)         & 0.2 \\
                            &                     & $\nu_{eff}^{k-\epsilon}$ & 0.05 (0.41) & -0.23 (0)         & 0.25 \\
                            &                     & $\nu_{eff}^{lsq}$        & 0.44 (0.41) & -0.01 (0)         & 0.24 \\
\hline
\multirow{3}{*}{C1 URANS}   &  \ldelim\{{3}{11pt} & $\nu_{mol}$       & 0.53 (0.41) & \textcolor{black}{-0.0375} (0)         & 0.15 \\
                            &                     & $\nu_{eff}^{k-\epsilon}$ & 0.06 (0.41) & -0.15   (0)         & 0.21 \\
                            &                     & $\nu_{eff}^{lsq}$        & 0.44 (0.41) & -0.0009 (0)         & 0.25 \\
\bottomrule
\end{tabular}
\caption{The Strouhal number predictions of the global mode based on PIV and URANS (C1) compared to the measured value which is given in brackets. The predicted growth rate $\Im(\omega_{g})$ is compared to the theoretical on the limit cycle of the global mode (given in brackets). $\omega_{g}$ is normalized via $\omega_{g}\mathrm{D}/v_{bulk}$. The wavemaker location is indicated by $\ts{x}{WM}$.  \label{tab:freqSelCFD}} 
\end{table}

The growth rate predictions are indicated by $\Im(\omega_{g})$ in \cref{tab:freqSelCFD}. The growth rate predicted with $\nu_{mol}$ and $\nu_{t}^{lsq}$ are of the order of magnitude of \textcolor{black}{$10e-2$} and $10e-4$, respectively. This is an excellent agreement with the theoretical value. As for the experimental results, the growth rate is predicted strongly negative with $\nu_{t}^{k-\epsilon}$. The fact that the $k-\epsilon$ based eddy viscosity performs equally poor for the URANS and all PIV cases further suggests that a lack of spatial resolution is not the dominating problem in the PIV measurements, but the focus on the normal stresses.
\section{Summary and discussion}
We investigated the accuracy of local linear stability analysis in predicting the frequency and growth rate of the global mode in \textit{turbulent} swirling jets undergoing vortex breakdown. Based on an analysis of the time-mean flow, we show that the frequency and growth rate can be predicted in good agreement with experimental results, if the non-coherent turbulence is included via an eddy viscosity. The error is below 7\% for the frequency prediction and the predicted growth rates are nearly identical to the zero growth rate of the global mode on the limit cycle.

We have demonstrated how the eddy viscosity can be obtained from PIV data sets via a least-square fit to the Reynolds stress tensor and via a computation of the turbulent kinetic energy $k$ and the turbulent dissipation rate $\epsilon$. The eddy viscosity obtained from the least-square approach takes into account the full Reynolds stress tensor and significantly improves the predictive capability of the stability analysis. Taking into account the eddy viscosity obtained from $k$ and $\epsilon$ deteriorated the predictions of the stability analysis, with a deviation from experimental measured frequencies by up to almost 90\%. An analysis based solely on the molecular viscosity deviated by at most 28\% and often even performed significantly better. We find that the eddy viscosity obtained from $k$ and $\epsilon$ focuses too strongly on the normal stress components. These are particularly strong at the upstream end of the recirculation bubble, where the global mode wavemaker is located. \textcolor{black}{Notwithstanding the present results, the $k-\epsilon$ approach may be a convenient and accurate alternative in flows with a less pronounced anisotropy. }

We further investigated whether URANS computations can supplement experimental techniques in the investigation of the dominant coherent structure in swirling jets undergoing vortex breakdown. Our findings reemphasize the observations made from the analysis of the PIV data. Only second moment closure that takes into account the full Reynolds stress tensor yields simulations of the flow field that are in qualitative and quantitative agreement with experimental findings. Any two-equation model that was attempted in this study yielded results that were in qualitative disagreement with experiments. Linear stability analysis based on time-averaged quantities of the second moment closure simulation yielded predictions of the global mode frequency that agreed to within 7\% with experiments and a global mode growth rate of zero ($10e-4$), which is consistent with a global mode oscillating on a limit cycle.

\textcolor{black}{Interestingly, the LSA equations can be closed with an appropriate eddy viscosity and yield accurate predictions, whereas the attempt to use the Boussinesq hypothesis in the prediction of the mean flow fails. Presumably, turbulence closure is less delicate for linear stability analysis since the mean flow, on which the analysis is based on, contains information about the turbulent fluctuations. Particularly inviscid instabilities as those considered here, should be largely determined by the mean flow field. Moreover, an isotropic eddy viscosity fitted to the full Reynolds stress tensor in a least squares sense represents the best possible approximation of the anisotropic eddy viscosity tensor, and it should always be superior to any modeling attempt.}  

How good is the present agreement of below 10\% between local LSA predictions and measurement in the context of other methods? Juniper et al. \cite{Juniper.2011} point out that the beauty of a local analysis lies in its ability to identify physically relevant mechanisms and that accurate results are best obtained by a global LSA. Paredes et al. \cite{Paredes.2016} conducted local and global stability analysis of the time-mean flows of different flame configurations present in a swirl stabilized combustor. They based their analysis on measured velocity data and derived the turbulence quantities a posteriori from the measured data. An exact number is not provided, but these authors state that both, the local and global analysis, are in reasonable agreement with the measured frequency of the global mode. They point out that the elliptic operators of global stability analysis are immensely sensitive to boundary conditions that had to be placed close to the wavemaker of the global mode, because of the limited optical access of the measurement setup. 

The global LSA of Crouch et al. \cite{Crouch.2007} considered the onset of flow unsteadiness over an airfoil by linearizing the URANS equations, including a turbulence model. They also provide an analysis of the laminar vortex shedding in the wake of a circular cylinder. For the latter case, global LSA predicts the shedding frequency within a deviation of $1e-3$ compared to the unsteady simulation of the full Navier-Stokes equation. The growth rate is predicted identical to the growth rate derived from the simulation. No exact numbers are given for the former case, but the authors state that the agreement between measurement and global LSA predictions is very good. Judging from the graphs they present, a small single digit percental deviation is conceivable.

Meliga et al. \cite{Meliga.2012b} base their global LSA on linearized RANS equations and the time-mean flow behind a D-shaped cylinder. They find that the frequency of the global mode predicted by global LSA deviates by around 9\% from the value obtained from the unsteady simulation. The authors argue in line with Sipp \& Lebedev \cite{Sipp.2007} that the match between global LSA prediction and simulation is not exact, because of a strong resonance of the global mode with its first harmonic that may render the analysis of the time-mean flow questionable. The growth rate of the global mode is predicted significantly different from zero for the same reason.

These studies indicate that global LSA may be capable of predicting the growth rate and frequency of the global mode (almost) exactly. For global LSA to perform significantly better than local LSA, appropriate boundary conditions have to be placed sufficiently far away from the wavemaker of the global mode. However, obtaining the flow field far upstream and downstream of the global mode may not be possible in an experimental setting. Under these circumstances, a deviation of roughly 10\% between measurement and prediction is a very good value.           
\section{Conclusion}
Local linear stability analysis based on time-mean velocity fields predicts the frequency and growth rate of the global mode in highly turbulent swirling jets undergoing vortex breakdown in good agreement with experiments \textcolor{black}{and theory}, if fine scale turbulence is included in the analysis via an appropriate turbulence model. The turbulence model for the stability analysis predictions has to be chosen with care and needs to take into account the turbulence structure of the flow at hand. For the flows considered, only a model of the full Reynolds stress tensor improved the predictions, while a model based on $k$ and $\epsilon$ deteriorated the accuracy of the predicted growth rate and frequency of the global mode. These observations are expected to extend to global stability analysis and also to other flows with a complex turbulence structure.

The observations based on analyses of measured data are corroborated by URANS simulations. Only a second moment turbulence closure model was found to predict the flow in close agreement with the measurements. 
Consistent URANS predictions might be an attractive basis for global stability analyses, as they provide velocity data sufficiently far upstream of the region of interest and thus ease the definition of appropriate boundary conditions.   
    
\section*{Acknowledgement}
The financial support of the German Research Foundation (DFG) under project PA 920/29-1 is acknowledged.
\appendix
\gdef\thesection{\Alph{section}} 
\makeatletter
\renewcommand\@seccntformat[1]{Appendix \csname the#1\endcsname.\hspace{0.5em}}
\makeatother

\section{Validation of the URANS computation}
\label{sec:cfdVali}
\begin{figure}
\centering
\includegraphics{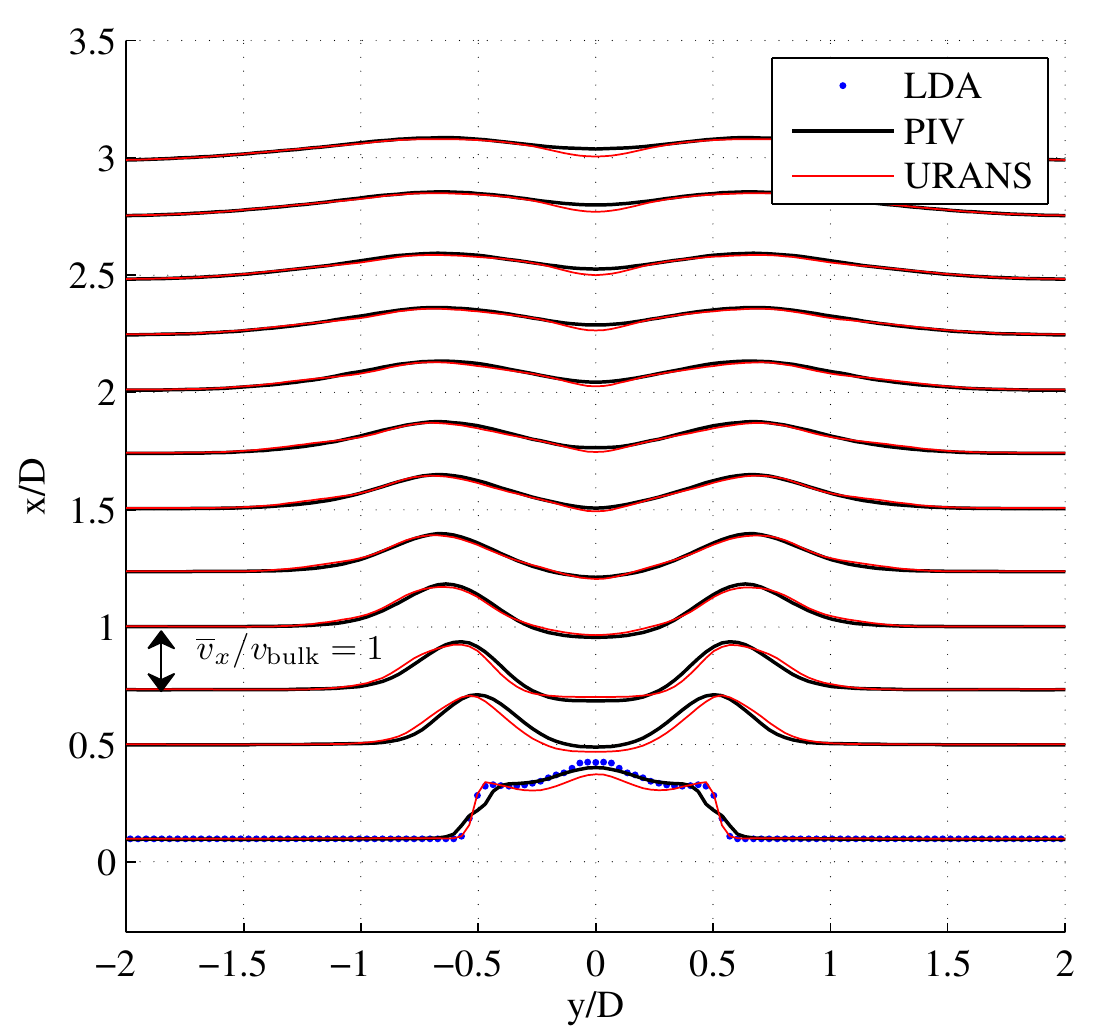}
\caption{Axial velocity profiles at various axial stations.}
\label{fig:axialVeloLine} 
\end{figure}

This appendix discusses the time-mean and the dynamic properties of configuration C1, obtained from PIV and from URANS. \textcolor{black}{The URANS computations were obtained with the Reynolds stress model of Launder et al. \cite{Launder.1975}}. The good correspondence between the axial velocity fields that is evident in \cref{fig:meanVelo2} is further substantiated in \cref{fig:axialVeloLine}. It shows axial velocity profiles at different axial stations. For clarity of presentation, velocity profiles in the immediate near field are only shown at $x/\mathrm{D} = 0.1$. At this position, additional Laser-Doppler-Anemometry (LDA) measurements are available for comparison. The distance from the nozzle lip was necessary to accommodate the LDA beam spacing. An average of about 330000 samples were collected at each measurement point for the axial component.

The URANS velocity profile at $x/\mathrm{D} = 0.1$ agrees well with the measured profiles. The overshoot on the jet axis is smaller for the URANS profile than for those obtained from experiments. The mass flow rate in the URANS computation was 2\% smaller than that of the experiments, which is likely related to the estimation of the inflow conditions from another simulation. Given the overall good agreement of the time-mean velocities and the dynamics of the global mode with the experiment (\cref{fig:globalModeCFDvsExp}), this discrepancy is not detrimental to the quality of the simulation. URANS predictions and LDA measurements both indicate a steep gradient in the outer shear layer. PIV significantly overestimates the shear layer width in this area. The lacking resolution of strong gradients is a common problem of the correlation based PIV evaluation. A typical size of the interrogation windows size of 32x32 pixel may result in only very few velocity vectors across a strong gradient, resulting in insufficient resolution. The agreement of PIV and URANS increases with downstream distance from the nozzle. Particularly, the shear layer widths match well. The URANS computations overestimate the depth of the wake for $x/\mathrm{D} > 1.5$.      

\begin{figure}
\centering
\includegraphics{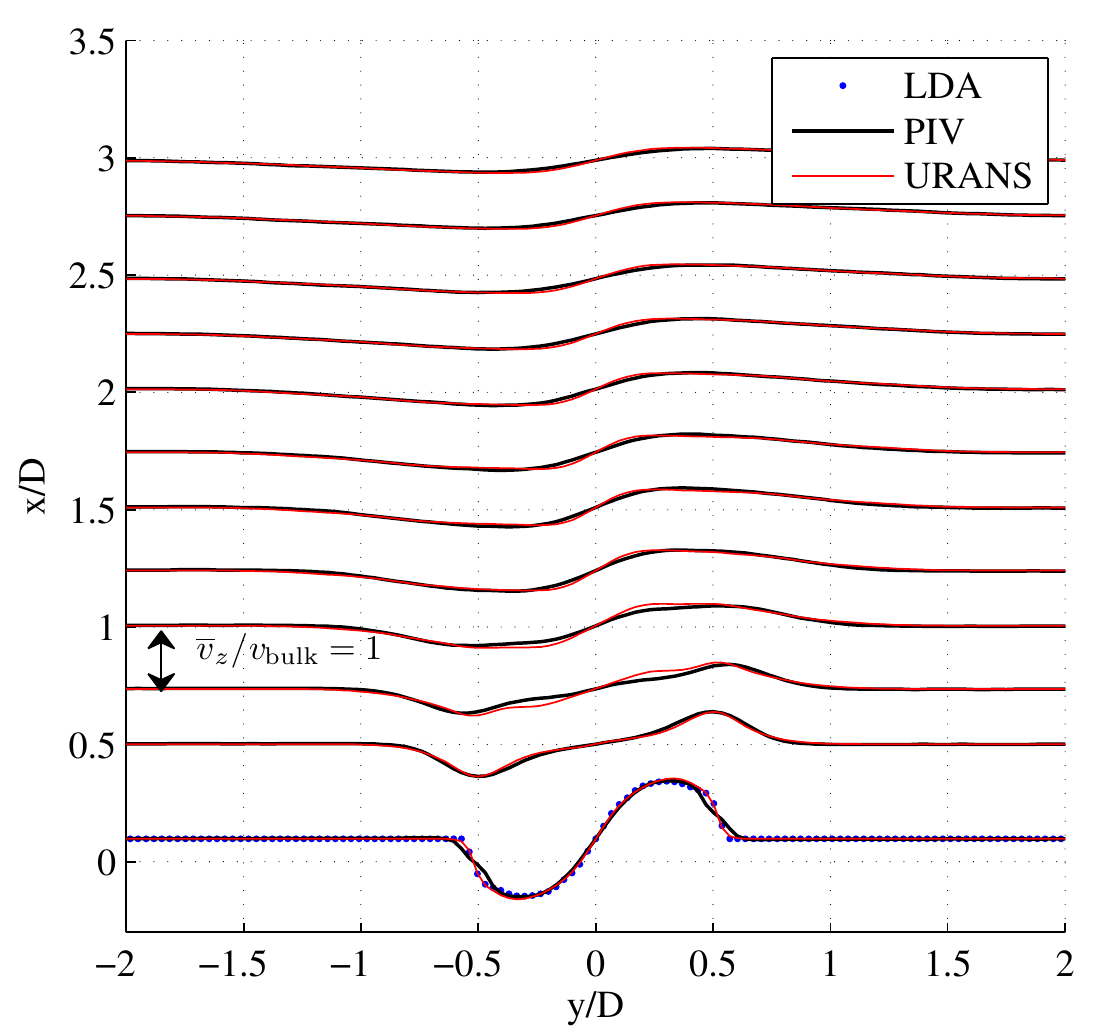}
\caption{Azimuthal velocity profiles at various axial stations.}
\label{fig:azimVeloLine} 
\end{figure}

\begin{figure}
\centering
\includegraphics{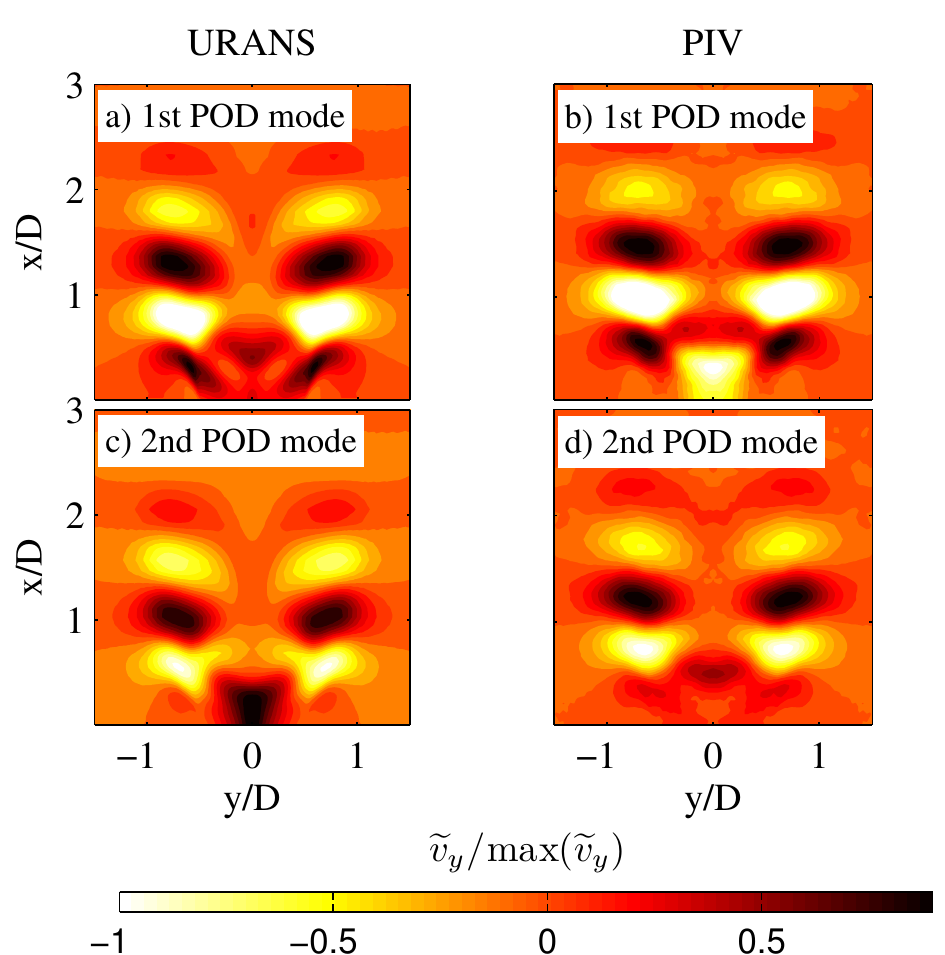}
\caption{Coherent cross stream velocity component of the global mode POD mode pair at arbitrary phase angle. a) and c) URANS. b) and d): PIV}
\label{fig:globalModeCFDvsExp} 
\end{figure}

A comparison between the experimentally and numerically obtained azimuthal velocity profiles is provided in \cref{fig:azimVeloLine}. As for \cref{fig:axialVeloLine}, LDA measurements are available at $x/\mathrm{D} = 0.1$. The number of samples were around 10000 in this case. A good agreement between PIV and URANS is evident throughout the domain. The correspondence between LDA and URANS in the vicinity of the nozzle is good and validates the URANS results.

\begin{figure}
\centering
\includegraphics{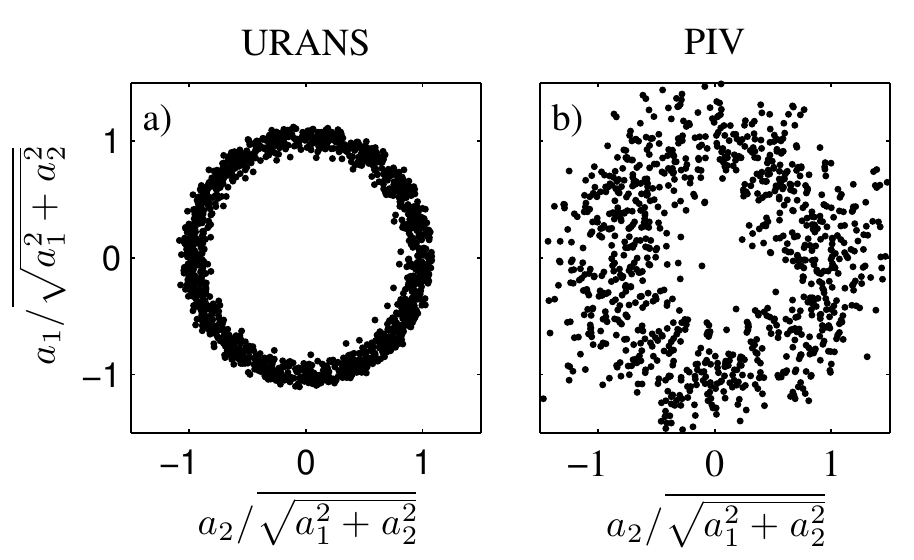}
\caption{Phase portrait of the POD modal amplitudes representing the limit cycle of the global mode from URANS (a) and PIV (b).}
\label{fig:modeCoeffCFDvsExp} 
\end{figure}

The dynamics of the global mode are compared next. The coherent cross stream velocity component of the dominant POD mode pair that represents the global mode is shown in \cref{fig:globalModeCFDvsExp}. This figure demonstrates that the URANS computations are capable of reproducing the dominant dynamics of the flow and of providing the correct spatial shape of the global mode structure. The periodicity of the POD modes is evident from the phase portrait of the model coefficients shown in \cref{fig:modeCoeffCFDvsExp}. The URANS computations accurately reproduce the pattern of a periodic structure that orbits on a limit cycle. The presence of ("resolved'') fine scale turbulence in the experimental data leads to fluctuations in the modal amplitude of the global mode (\cref{fig:modeCoeffCFDvsExp} b)). This is reflected in the fluctuations around a mean value present in \cref{fig:modeCoeffCFDvsExp} b). In any event, both phase portraits indicate the same periodicity. The simulation predicts the frequency of the global mode with 45 Hz, which is in good agreement with the 47 Hz derived from experiments.

Grid independence is established by repeating the simulation on a hexahedral mesh with circa 1.8 million cells. The inflow velocity is found to differ by at most 5\% and the frequency of the global mode by 3\% compared to the coarser mesh.   

\bibliographystyle{elsarticle-num} 
\bibliography{biblio}

\begin{thebibliography}{10}
\expandafter\ifx\csname url\endcsname\relax
  \def\url#1{\texttt{#1}}\fi
\expandafter\ifx\csname urlprefix\endcsname\relax\def\urlprefix{URL }\fi
\expandafter\ifx\csname href\endcsname\relax
  \def\href#1#2{#2} \def\path#1{#1}\fi

\bibitem{Lugo.2014}
V.~Manti{\v{c}}-Lugo, C.~Arratia, F.~Gallaire, Self-consistent mean flow
  description of the nonlinear saturation of the vortex shedding in the
  cylinder wake, Physical Review Letters 113~(8).

\bibitem{Noack.2003}
B.~R. Noack, K.~Afanasiev, M.~Morzynski, G.~Tadmor, F.~Thiele, A hierarchy of
  low-dimensional models for the transient and post-transient cylinder wake,
  Journal of Fluid Mechanics 497 (2003) 335--363.

\bibitem{Barkley.2006}
D.~Barkley, Linear analysis of the cylinder wake mean flow, Europhys. Lett
  75~(5) (2006) 750--756.

\bibitem{Gaster.1985}
M.~Gaster, E.~Kit, I.~Wygnanski, Large-scale structures in a forced turbulent
  mixing layer, Journal of Fluid Mechanics 150 (1985) 23--39.

\bibitem{Cohen.1994}
J.~Cohen, B.~Marasli, V.~Levinski, The interaction between the mean flow and
  coherent structures in turbulent mixing layers, Journal of Fluid Mechanics
  260~(-1) (1994) 81.

\bibitem{Reau.2002b}
N.~Reau, A.~Tumin, On harmonic perturbations in a turbulent mixing layer,
  European Journal of Mechanics - B/Fluids 21~(2) (2002) 143--155.

\bibitem{Marasli.1994}
B.~Marasli, J.~Cohen, V.~Levinski, Mean flow distortion due to finite-amplitude
  instability waves in a plane turbulent wake, Physics of Fluids 6~(3) (1994)
  1315--1322.

\bibitem{Pier.2002}
B.~Pier, On the frequency selection of finite-amplitude vortex shedding in the
  cylinder wake, Journal of Fluid Mechanics 458 (2002) 407--417.

\bibitem{Hammond.1997}
D.~A. Hammond, L.~G. Redekopp, Global dynamics of symmetric and asymmetric
  wakes, Journal of Fluid Mechanics 331 (1997) 231--260.

\bibitem{LEONTINI.2010}
J.~S. Leontini, M.~C. Thompson, K.~Hourigan, A numerical study of global
  frequency selection in the time-mean wake of a circular cylinder, Journal of
  Fluid Mechanics 645 (2010) 435--446.

\bibitem{Khor.2008}
M.~Khor, J.~Sheridan, M.~C. Thompson, K.~Hourigan, Global frequency selection
  in the observed time-mean wakes of circular cylinders, Journal of Fluid
  Mechanics 601 (2008) 425--441.

\bibitem{Thiria.2007}
B.~Thiria, J.~E. Wesfreid, Stability properties of forced wakes, Journal of
  Fluid Mechanics 579 (2007) 137.

\bibitem{Thiria.2015}
B.~Thiria, G.~Bouchet, J.~E. Wesfreid, On the relation between linear stability
  analysis and mean flow properties in wakes., arXiv:1506.05948 (2015)
  [physics.flu--dyn].

\bibitem{Turton.2015}
S.~E. Turton, L.~S. Tuckerman, D.~Barkley, Prediction of frequencies in
  thermosolutal convection from mean flows, Physical Review E 91~(4).

\bibitem{Oberleithner.2011b}
K.~Oberleithner, M.~Sieber, C.~N. Nayeri, C.~O. Paschereit, C.~Petz, H.-C.
  Hege, B.~R. Noack, I.~Wygnanski, Three-dimensional coherent structures in a
  swirling jet undergoing vortex breakdown: stability analysis and empirical
  mode construction, Journal of Fluid Mechanics 679 (2011) 383--414.

\bibitem{Oberleithner.2015c}
K.~Oberleithner, M.~St{\"o}hr, S.~H. Im, C.~M. Arndt, A.~M. Steinberg,
  Formation and flame-induced suppression of the precessing vortex core in a
  swirl combustor: Experiments and linear stability analysis, Combustion and
  Flame 162~(8) (2015) 3100--3114.

\bibitem{Oberleithner.2014b}
K.~Oberleithner, L.~Rukes, J.~Soria, Mean flow stability analysis of
  oscillating jet experiments, Journal of Fluid Mechanics 757 (2014) 1--32.

\bibitem{Paschereit.1995}
C.~O. Paschereit, I.~Wygnanski, H.~E. Fiedler, Experimental investigation of
  subharmonic resonance in an axisymmetric jet, Journal of Fluid Mechanics
  283~(-1) (1995) 365.

\bibitem{Sipp.2007}
D.~Sipp, A.~Lebedev, Global stability of base and mean flows: a general
  approach and its applications to cylinder and open cavity flows, Journal of
  Fluid Mechanics 593.

\bibitem{Terhaar.2015a}
S.~Terhaar, K.~Oberleithner, C.~O. Paschereit, Suppression and excitation of
  the precessing vortex core by acoustic velocity fluctuations: An experimental
  and analytical study, Combustion and Flame under review.

\bibitem{Reynolds.1972}
W.~C. Reynolds, A.~K. M.~F. Hussain, The mechanics of an organized wave in
  turbulent shear flow. part 3. theoretical models and comparisons with
  experiments, Journal of Fluid Mechanics 54~(02) (1972) 263--288.

\bibitem{Crouch.2007}
J.~Crouch, A.~Garbaruk, D.~Magidov, Predicting the onset of flow unsteadiness
  based on global instability, Journal of Computational Physics 224~(2) (2007)
  924--940.

\bibitem{Meliga.2012b}
P.~Meliga, G.~Pujals, E.~Serre, Sensitivity of 2-d turbulent flow past a
  d-shaped cylinder using global stability, Physics of Fluids 24~(6) (2012)
  61701.

\bibitem{Viola.2014}
F.~Viola, G.~V. Iungo, S.~Camarri, F.~Port{\'e}-Agel, F.~Gallaire, Prediction
  of the hub vortex instability in a wind turbine wake: stability analysis with
  eddy-viscosity models calibrated on wind tunnel data, Journal of Fluid
  Mechanics 750.

\bibitem{Kitsios.2010b}
V.~Kitsios, L.~Cordier, J.-P. Bonnet, A.~Ooi, J.~Soria, Development of a
  nonlinear eddy-viscosity closure for the triple-decomposition stability
  analysis of a turbulent channel, Journal of Fluid Mechanics 664 (2010)
  74--107.

\bibitem{Kitsios.2011}
V.~Kitsios, L.~Cordier, J.-P. Bonnet, A.~Ooi, J.~Soria, On the coherent
  structures and stability properties of a leading-edge separated aerofoil with
  turbulent recirculation, Journal of Fluid Mechanics 683 (2011) 395--416.

\bibitem{Syred.1997}
N.~Syred, W.~Fick, T.~O'Doherty, A.~J. Griffiths, The effect of the precessing
  vortex core on combustion in a swirl burner, Combustion Science and
  Technology 125~(1-6) (1997) 139--157.

\bibitem{Stohr.2012}
M.~St{\"o}hr, I.~Boxx, C.~D. Carter, W.~Meier, Experimental study of
  vortex-flame interaction in a gas turbine model combustor, Combustion and
  Flame 159~(8) (2012) 2636--2649.

\bibitem{Rukes.2015}
L.~Rukes, M.~Sieber, C.~O. Paschereit, K.~Oberleithner, Effect of initial
  vortex core size on the coherent structures in the swirling jet near field,
  Experiments in Fluids 56~(10).

\bibitem{Chigier.1965}
N.~A. Chigier, A.~Chervinsky, Experimental and theoretical study of turbulent
  swirling jets issuing from a round orifice, {Technion, Israel Institute of
  Technology, Dept. of Aeronautical Engineering}, Haifa and Israel, 1965.

\bibitem{Panda.1994}
J.~Panda, D.~K. McLaughlin, Experiments on the instabilities of a swirling jet,
  Physics of Fluids 6~(1) (1994) 263--276.

\bibitem{Billant.1998}
P.~Billant, J.-M. Chomaz, P.~Huerre, Experimental study of vortex breakdown in
  swirling jets, Journal of Fluid Mechanics 376 (1998) 183--219.

\bibitem{Ruith.2003b}
M.~R. Ruith, P.~Chen, E.~Meiburg, T.~Maxworthy, Three-dimensional vortex
  breakdown in swirling jets and wakes: direct numerical simulation, Journal of
  Fluid Mechanics 486 (2003) 331--378.

\bibitem{Liang.2005}
H.~Liang, T.~Maxworthy, An experimental investigation of swirling jets, Journal
  of Fluid Mechanics 525 (2005) 115--159.

\bibitem{Oberleithner.2012}
K.~Oberleithner, R.~Seele, C.~O. Paschereit, I.~Wygnanski, The formation of
  turbulent vortex breakdown: intermittency, criticality, and global
  instability, AIAA Journal 50~(7) (2012) 1437--1452.

\bibitem{Semaan.2013}
R.~Semaan, J.~W. Naughton, Three-component laser-doppler-anemometry
  measurements in turbulent swirling jets, AIAA Journal 51~(9) (2013)
  2098--2113.

\bibitem{Selle.2004}
L.~Selle, G.~Lartigue, T.~Poinsot, R.~Koch, K.-U. Schildmacher, W.~Krebs,
  B.~Prade, P.~Kaufmann, D.~Veynante, Compressible large eddy simulation of
  turbulent combustion in complex geometry on unstructured meshes, Combustion
  and Flame 137~(4) (2004) 489--505.

\bibitem{Willert.1991}
C.~Willert, M.~Gharib, Digital particle image velocimetry, Experiments in
  Fluids 10~(4) (1991) 181--193.

\bibitem{Scarano.2002}
F.~Scarano, Iterative image deformation methods in piv, Measurement Science and
  Technology 13~(1) (2002) R1.

\bibitem{Soloff.1997}
S.~M. Soloff, R.~J. Adrian, Z.-C. Liu, Distortion compensation for generalized
  stereoscopic particle image velocimetry, Measurement Science and Technology
  8~(12) (1997) 1441.

\bibitem{Weller.1998}
H.~G. Weller, G.~Tabor, H.~Jasak, C.~Fureby, A tensorial approach to
  computational continuum mechanics using object-oriented techniques, Computers
  in physics 12~(6) (1998) 620--631.

\bibitem{Issa.1986}
R.~Issa, Solution of the implicitly discretised fluid flow equations by
  operator-splitting, Journal of Computational Physics 62~(1) (1986) 40--65.

\bibitem{Shih.1997}
T.-H. Shih, J.~Zhu, W.~Liou, K.-H. Chen, N.-S. Liu, Modelling of turbulent
  swirling flows, NASA Technical Memorandum 113112.

\bibitem{Guo.2001}
B.~Guo, T.~A.~G. Langrish, D.~F. Fletcher, Simulation of turbulent swirl flow
  in an axisymmetric sudden expansion, AIAA Journal 39~(1) (2001) 96--102.

\bibitem{Wegner.2004}
B.~Wegner, A.~Maltsev, C.~Schneider, A.~Sadiki, A.~Dreizler, J.~Janicka,
  Assessment of unsteady rans in predicting swirl flow instability based on les
  and experiments, International Journal of Heat and Fluid Flow 25~(3) (2004)
  528--536.

\bibitem{Jochmann.2006}
P.~Jochmann, A.~Sinigersky, M.~Hehle, O.~Sch{\"a}fer, R.~Koch, H.-J. Bauer,
  Numerical simulation of a precessing vortex breakdown, International Journal
  of Heat and Fluid Flow 27~(2) (2006) 192--203.

\bibitem{Jones.1972}
W.~Jones, B.~Launder, The prediction of laminarization with a two-equation
  model of turbulence, International Journal of Heat and Mass Transfer 15~(2)
  (1972) 301--314.

\bibitem{Shih.1995}
T.-H. Shih, W.~W. Liou, A.~Shabbir, Z.~Yang, J.~Zhu, A new k-$\epsilon$ eddy
  viscosity model for high reynolds number turbulent flows, Computers {\&}
  Fluids 24~(3) (1995) 227--238.

\bibitem{Yakhot.1992}
V.~Yakhot, S.~A. Orszag, S.~Thangam, T.~B. Gatski, C.~G. Speziale, Development
  of turbulence models for shear flows by a double expansion technique, Physics
  of Fluids A 4~(7) (1992) 1510--1520.

\bibitem{Menter.1994}
F.~R. Menter, Two-equation eddy-viscosity turbulence models for engineering
  applications, AIAA Journal 32~(8) (1994) 1598--1605.

\bibitem{Shih.1993}
T.-H. Shih, J.~Zhu, J.~L. Lumley, A realizable reynolds stress algebraic
  equation model, NASA Technical Memorandum 105993.

\bibitem{Launder.1975}
B.~E. Launder, G.~J. Reece, W.~Rodi, Progress in the development of a
  reynolds-stress turbulence closure, Journal of Fluid Mechanics 68~(03) (1975)
  537.

\bibitem{Holmes.1998b}
P.~Holmes, J.~Lumley, G.~Berkooz, Turbulence, Coherent Structures, Dynamical
  Systems and Symmetry, Cambridge Monographs on Mechanics, {Cambridge
  University Press}, Cambridge [England], 1998.

\bibitem{Khorrami.1989}
M.~R. Khorrami, M.~R. Malik, R.~L. Ash, Application of spectral collocation
  techniques to the stability of swirling flow, Journal of Computational
  Physics 81 (1989) 206--229.

\bibitem{Boyd.2001}
J.~P. Boyd, Chebyshev and Fourier Spectral Methods, {Dover Publications,
  Mineola, New York}, 2001.

\bibitem{Trefethen.2000}
L.~N. Trefethen, Spectral Methods in Matlab, {Society for Industrial and
  Applied Mathematics}, 2000.

\bibitem{Fornberg.1998}
B.~Fornberg, A practical guide to pseudospectral methods, Cambridge monographs
  on applied and computational mathematics, {Cambridge University Press},
  Cambridge, 1998.

\bibitem{Huerre.1990}
P.~Huerre, P.~A. Monkewitz, Local and global instabilities in spatially
  developing flows, Annual Review of Fluid Mechanics 22 (1990) 473--537.

\bibitem{Batchelor.2003}
G.~K. Batchelor, H.~K. Moffatt, M.~G. Worster, Perspectives in fluid dynamics:
  A collective introduction to current research, 1st Edition, {Cambridge
  University Press}, Cambridge, 2003, {\copyright}2000.

\bibitem{Chomaz.1991}
J.~M. Chomaz, P.~Huerre, L.~G. Redekopp, A frequency selection criterion in
  spatially developing flows, Stud. Appl. Math 84 (1991) 119--144.

\bibitem{Juniper.2011}
M.~P. Juniper, O.~Tammisola, F.~Lundell, The local and global stability of
  confined planar wakes at intermediate reynolds number, Journal of Fluid
  Mechanics 686 (2011) 218--238.

\bibitem{Ivanova.2012}
E.~Ivanova, Numerical simulations of turbulent mixing in complex flows, Ph.D.
  thesis, Stuttgart [Germany] (2012).

\bibitem{Allmaras.2012}
S.~Allmaras, F.~Johnson, P.~Spalart, Modifications and clarifications for the
  implementation of the spalart-allmaras turbulence model, in: Seventh
  International Conference on Computational Fluid Dynamics, 2012, 2012, pp.
  1--11.

\bibitem{Ilinca.1996}
F.~Ilinca, D.~Pelletier, A.~Garon, Positivity preserving formulations for
  adaptive solution of two-equation models of turbulence, in: Fluid Dynamics
  Conference, Fluid Dynamics and Co-located Conferences, {American Institute of
  Aeronautics and Astronautics}, 1996.
\newblock \href {http://dx.doi.org/10.2514/6.1996-2056}
  {\path{doi:10.2514/6.1996-2056}}.

\bibitem{Kraichnan.1976}
R.~H. Kraichnan, Eddy viscosity in two and three dimensions, J. Atmos. Sci.
  33~(8) (1976) 1521--1536.

\bibitem{Shah.1995}
K.~Shah, J.~Ferziger, A new non-eddy viscosity subgrid-scale model and its
  application to channel flow, Center for Turbulence Research Annual Research
  Briefs.

\bibitem{Rodi.1997}
W.~Rodi, J.~H. Ferziger, M.~Breuer, M.~Pourquiee, Status of large eddy
  simulation: Results of a workshop, Journal of Fluids Engineering 119~(2)
  (1997) 248.

\bibitem{Sheng.2000}
J.~Sheng, H.~Meng, R.~O. Fox, A large eddy piv method for turbulence
  dissipation rate estimation, Chemical Engineering Science 55~(20) (2000)
  4423--4434.

\bibitem{Smagorinsky.1963}
J.~Smagorinsky, Generall circulation experiments with the primitive equations,
  Mon. Wea. Rev. 91~(3) (1963) 99--164.

\bibitem{Delafosse.2011}
A.~Delafosse, M.-L. Collignon, M.~Crine, D.~Toye, Estimation of the turbulent
  kinetic energy dissipation rate from 2d-piv measurements in a vessel stirred
  by an axial mixel ttp impeller, Chemical Engineering Science 66~(8) (2011)
  1728--1737.

\bibitem{Tanaka.2007}
T.~Tanaka, J.~K. Eaton, A correction method for measuring turbulence kinetic
  energy dissipation rate by piv, Experiments in Fluids 42~(6) (2007) 893--902.

\bibitem{Saarenrinne.2000}
P.~Saarenrinne, M.~Piirto, Turbulent kinetic energy dissipation rate estimation
  from piv velocity vector fields, Experiments in Fluids 29~(7) (2000)
  S300--S307.

\bibitem{Rukes.2015b}
L.~Rukes, M.~Sieber, C.~O. Paschereit, K.~Oberleithner, Theoretical and
  experimental investigation of the helical global mode in non-isothermal open
  swirling jets, Physics of Fluids submitted.

\bibitem{Paredes.2016}
P.~Paredes, S.~Terhaar, K.~Oberleithner, V.~Theofilis, C.~{Oliver Paschereit},
  Global and local hydrodynamic stability analysis as a tool for combustor
  dynamics modeling, Journal of Engineering for Gas Turbines and Power 138~(2)
  (2016) 21504.

\end{thebibliography}

\end{document}